\documentclass[superscriptaddress,amsmath,amssymb]{revtex4}
\usepackage{graphicx}
\usepackage{dcolumn}
\usepackage{bm}
\usepackage{latexsym}
\begin{document} 
\title{Template-directed biopolymerization: tape-copying Turing machines{\footnote{Contribution in the Alan Turing year 2012}}} 
\author{Ajeet K. Sharma}
\affiliation{Department of Physics, Indian Institute of Technology, 
        Kanpur 208016, India} 
\author{Debashish Chowdhury}
\affiliation{Department of Physics, Indian Institute of Technology, 
        Kanpur 208016, India} 
\date{\today}
\begin{abstract} 
DNA, RNA and proteins are among the most important macromolecules in 
a living cell. These molecules are polymerized by molecular machines. 
These natural nano-machines polymerize such macromolecules, adding one 
monomer at a time, using another linear polymer as the corresponding 
template. The machine utilizes input chemical energy to move along 
the template which also serves as a track for the movements of the 
machine. In the Alan Turing year 2012, it is worth pointing out that 
these machines are ``tape-copying Turing machines''. We review the 
operational mechanisms of the polymerizer machines and their 
collective behavior from the perspective of statistical physics, 
emphasizing their common features in spite of the crucial differences 
in their biological functions. We also draw attention of the physics 
community to another class of modular machines that carry out a 
different type of template-directed polymerization. We hope this 
review will inspire new kinetic models for these modular machines. 
\end{abstract}

\maketitle

\newpage 

\begin{tabular}{|l|l|}
  \hline
  \multicolumn{2}{|c|}{Abbreviations} \\
  \hline
  A & Adenosine  \\
  T & Thymine  \\
  C & Cytosine \\
  G & Guanine  \\
  U & Uracil \\
  RNA & Ribonucleic acid \\
  DNA & Deoxyribonucleic acid \\
  mRNA & Messenger RNA \\
  rRNA & Ribosomal RNA\\
  tRNA & Transfer RNA \\
  NTP & Nucleotide triphosphate \\
  ATP & Adenosine triphosphate \\
  GTP & Guanosine triphosphate\\
  GDP & Guanosine diphosphate \\
  dNTP & Deoxyribonucleotide triphosphate \\
  DDRP  & DNA dependent RNA polymerase \\
  DDDP  & DNA dependent DNA polymerase \\
  RDRP & RNA dependent RNA polymerase\\
  RDDP & RNA dependent DNA polymerase \\
  RNAP &  RNA polymerase\\
  DNAP &  DNA polymerase \\
  dsDNA & Double stranded DNA \\
  ssDNA & Single stranded DNA \\ 
  $PP_{i}$ & Pyrophosphate \\ 
   TEC & Transcription elongation complex \\
   ASEP & Asymmetric simple exclusion process \\
   TASEP & Totally asymmetric simple exclusion process \\
   ADP & Adenosine diphosphate \\
  LTR & Long terminal repeats\\
  HIV & Human immunodeficiency virus\\
  AIDS & Acquired immunodeficiency syndrome \\
  RNaseH & Ribonuclease H \\  
  RT & Reverse transcriptase \\
  aa-tRNA & Aminoacyl tRNA \\ 
  aa-tRNA synth & Aminoacyl tRNA synthetase \\
  EF-Tu & Elongation factor thermo unstable \\
  EF-G & Elongation factor G\\
  DTD & Dwell time distribution \\
  L or Leu & Leucine \\
  Phe or F &  Phenylalanine \\
  UTR & Untranslated region \\
  NRPS & Nonribosomal peptide synthetase \\
  PKS & Polyketide synthase \\
  FAS & Fatty acid synthase\\
  KS & Ketosynthase \\
  AT & Acyl transferase \\  
  ACP & Acyl carrier protein \\
  KR & Ketoreductase \\
  DH & Dehydratase\\
  ER & Enoyl reductase \\
  PCP-domain & Peptidyl carrier domain \\  
 \hline
\end{tabular}
\\
\begin{tabular}{|l|l|}
  \hline
 \multicolumn{2}{|c|}{Abbreviations} \\
  \hline
  C-domain & Condensation domain \\
  A-domain & Adenylation domain \\
  MT-domain & Methyltransferase domain \\
  E-domain & Epimerisation domain\\
  TE-domain & Thioesterase domain\\  
  \hline
\end{tabular}
\\
\begin{tabular}{|l|l|}
  \hline
  \multicolumn{2}{|c|}{Symbols} \\
  \hline
  $\ell$ & Covering length of a molecular machine.  \\
  $Q(\underline{i}|j)$ & Conditional probability that, given a RNAP at site $i$, there
is no RNAP at site $j$.
  \\
  $\rho$ & Number density of molecular machines on its one dimensional track. \\
  $\rho_{cov}$ & Coverage density of molecular machines on its one dimensional track. \\
  J & Overall rate of RNA synthesis. \\
  k & Replication speed. \\
  F & Force applied on the dsDNA.  \\
  $f_{dwell}$ & Dwell time distribution of the ribosome. \\
  $E_{c}$ & Closed finger configuration of DNAP. \\
  $E_{o}$ & Open finger configuration of DNAP. \\
  $D_{n}$ & Position of the catalytic site of DNAP, on its one dimensional track ($n_{th}$).\\
  oriC  & Initiation site in replication process. \\ 
  terC & Termination site in replication process. \\
  ${\cal P}_{n}$ & Probability of reaching the RNAP at termination site($n=N$) by incorporating a correct \\
   & nucleotide at   position $n$.\\
  $P_{m}(t)$ & Probability of finding the TEC at $m$(relative position of the RNAP catalytic site \\
   & with last incorporated nucleotide) at time $t$, having started at $m$ = 0 at time $t$ = 0.
 \\
  \hline
\end{tabular}

\newpage
\section{Introduction} 

Biological information is chemically encoded in the sequence of the 
species of the monomeric subunits of a class of linear polymers that 
play crucial roles in sustaining and propagating ``life''. Nature also 
designed wonderful machineries for polymerizing such macromolecules, 
step by step adding one monomer at each step, using another existing 
biopolymer as the corresponding template \cite{frank11}. In this review we summarize 
the recent progress in understanding the common features of the 
structural design of these machines and stochastic kinetics of the 
polymerization processes. 

In 1937, Alan Turing developed an abstract concept of a computing machine 
which was later named after him. On the occasion of the birth centenary 
of Alan Turing this year (2012), we emphasize the striking similarities 
between a Turing machine and the machines for template-directed 
polymerization of macromolecules of life \cite{shapiro06}. 

In a cell there are three different types of macromolecules, 
namely, polynucleotides, polypeptides and polysachharides,  
which perform wide range of important functions. 
The individual {\it monomeric residues} that form nucleic acids and
proteins are {\it nucleotides} and {\it amino acids}, respectively.
Both these types of macromolecules are {\it unbranched} polymers. 
De-oxyribo nucleic acid (DNA) and Ribonucleic acid (RNA) are 
polynucleotides whereas proteins are polypeptides. 
Nature uses 20 different species of amino acid subunits to make 
proteins; each amino acid species is denoted by three-letter symbols. 
In contrast, nature uses 4 different types of nucleotides, denoted 
by the one-letter symbols A, T, C and G, for making DNA. Similarly, 
4 types of nucleotides used for making RNA are A, U, C, and G. 
Discovering the genetic code \cite{erdmann11} 
that connects the 4-letter alphabet of the polynucleotides with the 
20-letter alphabet of the polypeptides was one of the greatest 
puzzle-solving exercise in molecular biology of the 20th century.
Messenger RNA (mRNA), ribosomal RNA (rRNA) and transfer RNA (tRNA) 
together form the group of ``core'' RNAs. 
It is also worth pointing out that only mRNA is translated whereas 
rRNA and tRNA form key components of the machinery that carries out 
translation.

In spite of the differences between their constituent monomers as well 
as in their primary, secondary and tertiary structures, nucleic acids 
and proteins share some common features in the birth and maturation. 
The main stages in the synthesis of polynucleotides by the polymerase
machines are common:
(a) {\it initiation}: Once the polymerase encounters a specific sequence
    on the template that acts as a chemically coded start signal, it
    initiates the synthesis of the product. This stage is completed when
    the nascent product becomes long enough to stabilize the macromolecular
    machine complex against dissociation from the template. \\
(b) {\it elongation}: During this stage, the nascent product gets
    elongated by the addition of nucleotides.  \\
(c) {\it termination}: Normally, the process of synthesis is terminated,
    and the newly polymerized full length product molecule is released,
    when the polymerase encounters the {\it terminator} (or, stop)
    sequence on the template. However, we shall consider, almost
    exclusively, the process of {\it elongation}.
Other common features of template-directed polymerization are as follows:\\
(i) Both nucleic acids and proteins are made from a limited number of
different species of monomeric building blocks. \\
(ii) The sequence of the monomeric subunits to be used for synthesis
are directed by the corresponding template. \\ 
(iii) The process goes through three phases, namely, {\it initiation}, 
{\it elongation} and {\it termination}.\\
(iv) During the elongation phase, the polymers are elongated, step-by-step, 
by successive addition of monomers, one at a time. \\
(v) For template-directed polymerization, the selection of the correct 
molecular species of subunit requires a mechanism of ``molecular 
recognition''. However, if this mechanism is not perfect, errors can 
occur. 
The typical probability of the errors in the final product is about $1$ 
(a) in $10^{3}$ polymerized amino acids, in case of protein synthesis 
\cite{ogle05,zaher09}, 
(b) in $10^4$ polymerized nucleotides in case of mRNA synthesis  
\cite{sydow09},   and 
(c) in $10^9$ polymerized nucleotides in case of replication of DNA 
\cite{kunkel00}. 
Purely thermodynamic discrimination of different species of nucleotide 
monomers cannot account for such high fidelity of polymerization. 
A normal living cell has mechanisms of ``kinetic proofreading'' and 
``editing'' so as to detect and correct errors. A theory that explains 
the physical origin of dissipation in computation \cite{bennett82}  
also provides insight into the need for energy expenditure in proofreading
processes during the transfer of genetic information 
\cite{bennett79,bennett08}.\\
(vi) The primary product of the synthesis, namely, polynucleotide or
polypeptide, often requires ``processing'' whereby the modified
product matures into functional nucleic acid or protein, respectively.\\

The free energy released by each event of the phosphate ester hydrolysis, 
that elongates the polynucleotide by one monomer, serves as the input 
energy for driving the mechanical movements of the corresponding 
polymerase by one step on its track. Moreover, as we'll discuss in 
detail later, GTP molecules are hydrolyzed during the process of 
polymerization of polypeptides. Therefore, the machines for template-
directed polymerization are also regarded as molecular motors; these 
use the template itself also as the track for their translocation. 

The molecular machines that polymerize polynucleotides are called 
{\it polymerase} whereas a {\it ribosome} polymerizes a polypeptide.
The genome of both prokaryotic and eukaryotic cells consist of DNA. 
However, many viruses use RNA as their genetic material. 
For their multiplication, viruses need not only to make copies of 
their genetic material, but also to manufacture the proteinous 
materials for constructing the capsids into which the freshly 
copied genomes have to be packaged. However, wide varieties of 
viruses do not have their own polymerases and none have their 
own ribosomes. Therefore, viruses exploit the machinery of their 
host for their own gene expression and genome replication. 

A systematic and unambiguous nomenclature for polymerases 
is based on the nature of the template and product polynucleotides 
\cite{choi12}: Both DNA-dependent RNA polymerase (DDRP) and 
DNA-dependent DNA polymerase (DDDP) use DNA as their templates; 
however, the former synthesizes a RNA molecule whereas the latter 
polymerizes a DNA. Similarly, RNA-dependent RNA polymerase (RDRP) 
and RNA-dependent DNA polymerase, both of which use RNA molecules 
as the templates, synthesize RNA and DNA molecules, respectively.   
The DDRP and DDDP that drive transcription and genome replication, 
respectively, of a cell are usually referred to as RNA polymerase 
(RNAP) and DNA polymerase (DNAP).

In this review we focus on the operational mechanisms of single 
machines engaged in template-directed polymerization in isolation 
as well as the collective phenomena caused by the interactions 
of the machines when involved simultaneously in the process. 

The main quantities of interest in the context of the mechanism of 
a single machine are: (a) the rate of synthesis of the macromolecules, 
and (b) the fraction of erroneous monomers incorporated in the final 
product. 
Although most of the works initially focussed on the average rates 
and average error, the fluctuations in these quantities is receiving 
more attention in recent years. 

In a living cell most often polymerases do not work in isolation.
A double stranded DNA serves as the track simultaneously for several 
polymerases. Therefore, discovering the ``traffic-rules'' 
\cite{lipowsky01,parmeggiani03,evans03,popkov03,lipowsky05,lipowsky06} 
for the polymerases is essential for understanding the coordination of 
transcription of different genes as well as that between transcription  
and replication. In this context, we explore the different types of 
possible ``binary collisions'' between two polymerases and the 
corresponding outcomes of the collisions. We also review the studies 
on the causes and consequences of traffic-like collective movements of many 
machines of the same type simultaneously on the same track in the same 
direction. Finally, we draw attention of the biophysics community to 
some modular machines that also carry out ``template-directed'' 
polymerization of a different type; quantitative modeling of these 
machines and their mechanisms from the perspective of physicists is 
desirable.

\section{Common structural features of polynucleotide polymerases}

A polymerase is expected to have binding sites for (a) the template 
strand, (b) the nascent polynucleotide strand, and (c) the NTP subunits.
It must have a mechanism to recognize and select the appropriate NTP 
directed by the template and a mechanism to catalyze the addition of 
the NTP thus selected to the growing polynucleotide. It would also 
be desirable to correct any error immediately before proceeding to the 
next step. A polymerase must be able to step forward by one nucleotide 
on its template without completely destabilizing the ternary complex 
consisting of the polymerase, the template and the product. Finally, 
it must have mechanisms for initiation and termination of the 
polymerization process. For several of these functions, particularly 
for initiation and termination, it requires assistance of other proteins. 

There are several common architectural features of all polynucleotide
polymerases. The shape of the polymerase has some resemblance with the
``cupped right hand'' of a normal human being; the three major domains
of it are identified with ``fingers'', ``palm'' and ``thumb''. There
are, of course, some crucial differences in the details of the
architectural designs of these machines which are essential for their
specific functions. The most obvious functional commonality between
these machines is that these add nucleotides, the monomeric subunits
of the nucleic acids, one by one following the template encoded in the
sequence of the nucleotides of the template. However, in spite of the 
gross architectural similarities between the polymerases in prokaryotic 
and eukaryotic cells, there are significant differences in the 
primary sequences of these machines.

\section{DDRP and transcription} 

\subsection{Single DDRP: speed and fidelity of transcription} 

\begin{figure}[ht]
\includegraphics[angle=0,width=0.6\columnwidth]{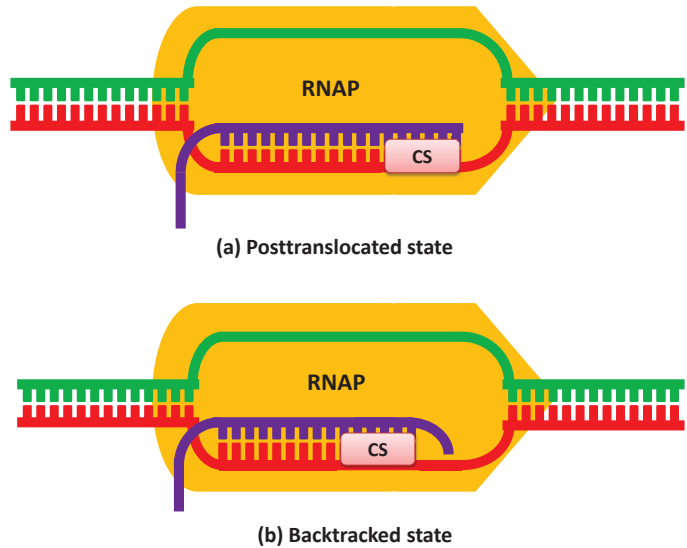}
\caption{Schematic representation of TEC in (a)post-translocated state (b) backtracked state}
\label{TEC}
\end{figure}

A common architectural feature of all DDRPs is the ``main internal
channel'' which can accomodate a DNA/RNA hybrid that is typically
$8$ to $9$ bp long. The NTP monomers enter through another pore-like
``entry channel'' while the nascent transcript emerges through the
``exit channel''. The formation of the bond between the newly arrived
NTP and the RNA chain takes place at a catalytically active site
located at the junction of the entry pore and the main channel.
In principle, during actual transcription, it may be necessary first
to unwind the DNA, at least locally, to get access to the nucleotide
sequence on a single-stranded DNA. Interestingly, the RNAP itself
exhibits helicase activity for this purpose.
A ``transcription-elongation complex'' (TEC), as shown schematically in
fig.\ref{TEC}, is formed by the RNAP, the dsDNA and the nascent RNA 
transcript. 

In the elongation stage, each cycle involves two main processes: 
polymerization and translocation. Polymerization elongates the RNA 
transcript by one nucleotide. During translocation, the RNAP moves 
forward by one nucleotide on the DNA template. The polymerase can 
fluctuate between two positions that are designated as ``pre-translocated'' 
and the ``post-translocated'' positions. In the absence of NTP, the 
RNAP executes an unbiased Brownian motion between these two positions.
However, an incoming NTP, upon binding, rectifies the fluctuations of 
the RNAP thereby biasing its movement in the forward direction. Thus 
NTP serves as a ``pawl'' in this Brownian ratchet mechanism
\cite{gelles98,bai07,sousa05,guo06,abbondanzieri05,cramer08,barnahum05}. 
Yu and Oster \cite{yu12} have developed a model that allow two parallel 
paths- one of these is based essentially on a Brownian ratchet mechanism 
whereas the other utilizes a power stroke.
Not all kinetic models explicitly assume either the power stroke or the
Brownian ratchet mechanism. Some purely kinetic models
\cite{tripathi08,yamada09,yamada10}
can be interpreted in either way because these assume movements of the
RNAP without explicitly explaining how these movements are caused by
the energy transduction mechanism.

Single molecule studies of DDRP have provided quantitative data on 
the force-velocity relation for these motors.
Calculation of the dwell time distribution \cite{tripathi09,yamada09} 
of RNAP is complicated because of the varieties of paused states. 
A RNAP can ``backtrack'' on its track, i.e., reverse translocate on its 
template by a few steps (see fig.\ref{TEC}). Paused states resulting 
from backtracking may be intrinsically different from the relatively 
shorter-lived paused states without backtracking \cite{landick06b}. 
The mechanisms of pausing and backtracking of RNAPs are still hotly 
debated 
\cite{shaevitz03,galburt07,bai04,landick09,dalal06}. 
Stochastic models have been developed for the kinetics of backtracking
\cite{voliotis08,depken09,xie08,xie09,xie12}.

Some of these models explicitly incorporate steps for proofreading
by either an isolated single RNAP \cite{voliotis09} or by individual
RNAP motors in a traffic of RNAPs \cite{sahoo11}.
In the model of nucleolytic proofreading developed by Voliotis et al. 
\cite{voliotis09} the integer index $n$ denotes the position of the 
last transcribed nucleotide on the template DNA. Another integer 
index $m$ denotes the position of the RNAP catalytic site with respect 
to $n$. For a fixed $n$, $P_{m}(t)$ is the probability of finding the TEC at $m$
at time $t$, having started at $m=0$ at time $t=0$.
Voliotis et al.\cite{voliotis09} defined ${\cal P}_{n}$
as the probabilities for reaching the termination site $n=N$, having 
incorporated the correct nucleotide at the position $n$ (and a similar 
probability for incorporating an incorrect nucleotide at $n$). 
Formulating a mathematical description, based on master equations for 
these probabilities, Voliotis et al.\cite{voliotis09} derived a site-wise 
detailed measure of the transcriptional error.

\subsection{Effects of RNAP-RNAP collision and RNAP traffic congestion} 

Two RNAPs can collide while transcribing either (i) the same gene, or 
(ii) two different genes. 
While transcribing the same gene simultaneously, the two RNAPs would 
move on the same DNA template strand and are co-directional. This 
situation is analogous to that of two vehicles in the same lane of a 
highway where both the vehicles are supposed to enter and exit the 
traffic at the same entry and exit points on this highway.
In such a co-directional collision, does the trailing polymerase get 
obstructed by the leading polymerase or does the leading polymerase 
get pushed from behind?
The leading RNAP may stall either because of backtracking or because 
of ``roadblocks'' created by other DNA-binding proteins. In both 
these situations, the co-directional trailing RNAP can rescue the 
stalled leading RNAP by ``pushing'' it forward from behind 
\cite{epshtein03a,epshtein03b}. 

Two distinct underlying mechanisms can manifest as ``pushing'' 
by the trailing RNAP \cite{galburt11}- in the first, the push exerted 
by the trailing RNAP on the leading stalled RNAP is a ``power stroke''; 
in the second, the leading stalled RNAP resumes transcription by 
thermal fluctuation just when the trailing one reaches it from behind 
thereby rectifying the backward movement of the leading RNAP by a 
``Brownian ratchet'' mechanism.
The elasticity of the TECs may give rise other possible outcomes of 
RNAP-RNAP collisions. For example, if the leading RNAP is ``obstructed'' 
by a sufficiently strong barrier, the trailing RNAP can backtrack after 
suffering collision with it \cite{saeki09}.

Next we consider the more complex situation where the two interacting 
RNAPs transcribe two different genes. 
The RNAP transcribing one gene can interfere with the {\it initiation}, 
or {\it elongation}, or {\it termination} of transcription of another 
neighbouring gene. The plausible scenarios of such ``transcriptional 
interference'' and the corresponding outcome of the RNAP-RNAP 
collisions have been studied quantitatively by a kinetic model  
\cite{sneppen05}. 
Simultaneous transcription of a gene by many DDRP motors can give rise 
to ``traffic congestion'' on the DNA track. Various aspects of this 
phenomenon, particularly, the effects of RNAP traffic congestion on  
on the average rate of transcription has been investigated by an 
extension of the asymmetric simple exclusion process (ASEP) 
\cite{tripathi08,klumpp08a,ohta11}. 
In a special case of ASEP, called totally asymmetric simple exclusion 
process (TASEP) particles cannot take backward step.

\begin{figure}[h]
\begin{center}
\includegraphics[angle=0,width=0.6\columnwidth]{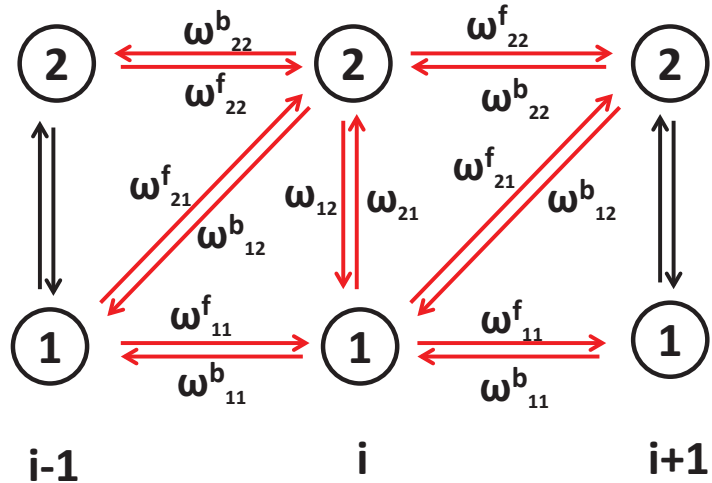}
\end{center}
\caption{Pictorial depiction of the two state model of RNAP (see the text for detials).}
\label{tranp}
\end{figure}

The kinetics of RNAP in a stochastic model \cite{tripathi08} is shown 
in figure \ref{tranp}. The rate of transition (i.e., transition 
probability per unit time) from state $j$ to the state $i$ is denoted 
by $\omega_{ij}$. In this two state model of RNAP \cite{tripathi08} 
integer index ``1'' is assigned to a state of RNAP, where no $PP_{i}$ 
is bound to it, while $PP_{i}$ bound RNAP is represented by symbol ``2''. 
Depending upon the different circumstances, polymerization step occurs 
with three different rates. If the NTP hydrolysis takes place (i) on RNAP 
(ii) in the solution while no $PP_{i}$ is bound to it and (iii) in the 
solution while $PP_{i}$ is bound with RNAP, then the polymerization step 
occurs with rate $\omega_{21}$, $\omega_{11}^{f}$ and $\omega_{22}^{f}$ 
respectively. Considering that all types of polymerization reactions are 
reversible, then the corresponding  backward transition rates are 
symbolized by $\omega_{12}^{b}$, $\omega_{11}^{b}$ and $\omega_{22}^{b}$, 
respectively. The release of  $PP_{i}$ ($2(i)\rightarrow 1(i)$) takes 
place with rate $\omega_{21}$ while its backward reaction occurs with 
rate $\omega_{12}$.\\ 

RNAP traffic moving on the DNA track resembles an ASEP where particles 
are replaced by rods of the length $\ell$ and often referred as 
$\ell$-ASEP. To indicate the position of RNAP on DNA track, we use the 
leftmost site among the $\ell$ successive sites covered by the RNAP. 
Let $Q(\underline{i}|j)$ be the conditional probability that, given a RNAP 
at site $i$ (expressed by the underbar), there is no RNAP at site $j$. 
Suppose $P_{\mu}(i,t)$ is the 
probability of finding the RNAP in $\mu_{th}$ chemical state at $i_{th}$ 
nucleotide at time t, then time evolution of these probabilities 
$P_{\mu}(i,t)$ are governed by master equations \cite{tripathi08} 
that incorporate the effective steric interaction felt by each RNAP. 

\begin{figure}[ht]
\begin{center}
\includegraphics[angle=0,width=0.6\columnwidth]{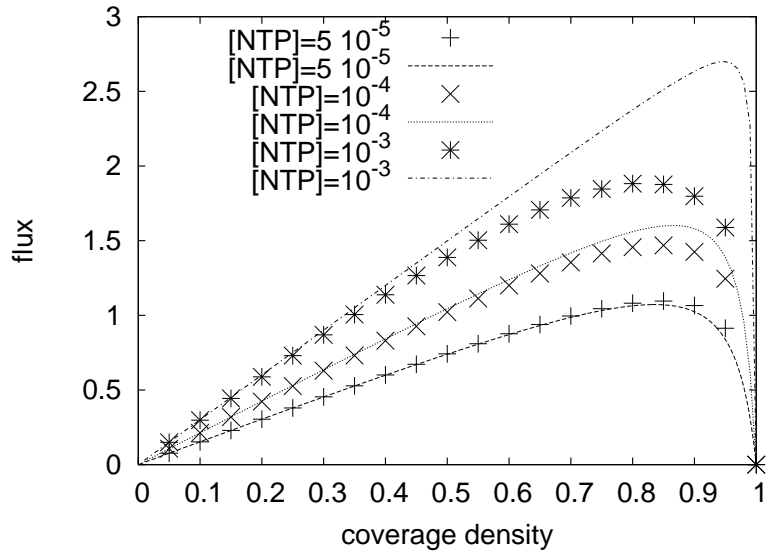}
\end{center}
\caption{RNAP flux is plotted for a few values of NTP concentration. Solid lines corresponds to the theoretical prediction whereas discrete data are obtained from the simulation. The values of other parameters are $\omega_{21}^{f}$=$10^{6}$[NTP]$s^{-1}$, $\omega_{11}^{f}$=46.6[NMP]$s^{-1}$, $\omega_{22}^{f}$=.31[NTP]$s^{-1}$, $\omega_{21}$=$10^{6}$[$PP_{i}$]$s^{-1}$, $\omega_{12}=31.4$ $s^{-1}$,$\omega_{12}^{b}=.21$ $s^{-1}$, $\omega_{11}^{b}=.9.4$ $s^{-1}$, $\omega_{22}^{b}=.063$ $s^{-1}$ and [$PP_{i}$]=1$\mu M$ (adapted from ref.\cite{tripathi08}). }
\label{tripathi3a}
\end{figure}

Under periodic boundary condition (PBC), the number density $\rho$ of 
the RNAPs is conserved. Solving the master equations under the steady 
state condition one can calculate the flux of the RNAPs analytically. 
Since in each polymerization step, mRNA is elongated by one nucleotide, 
RNAP flux also represents the net rate of mRNA synthesis. In figure 
\ref{tripathi3a}, flux J is plotted against the coverage density 
($\rho_{cov}=\rho \ell$) of RNAP, for a few different values of NTP 
concentration. Because of the extended size of the RNAP particles,  
the flux-density relation is asymmetric about $\rho=1/2$.

\begin{figure}[ht]
\includegraphics[angle=0,width=0.9\columnwidth]{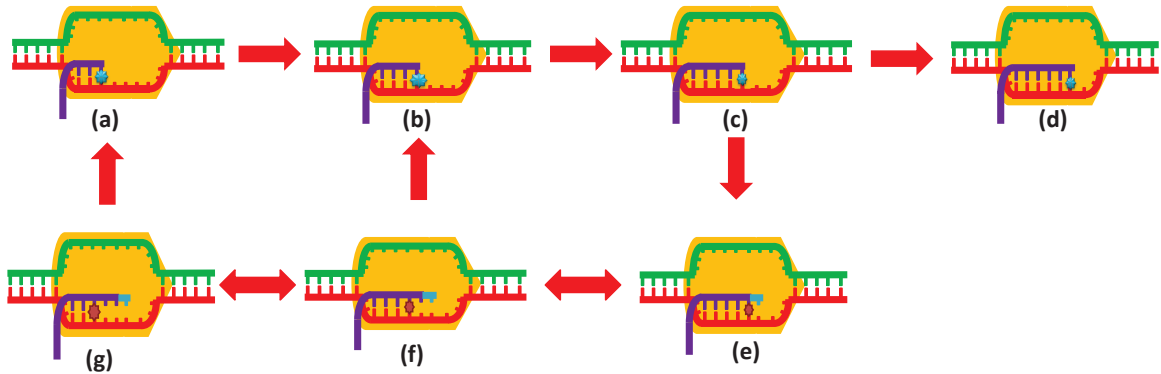}
\caption{Cartoon diagram of backtracking and kinetic proofreading during transcription. Active RNAP (state a) incorporates two successive correct nucleotides and reaches into state (c). After an incorrect incorporation it jumps in the error state (e), without any translocation. From error state(e), RNAP may backtrack to the previous site. In backtracked state RNAP has the equal probabibility for forward and backward motion. This backtracked state may also cleave the the erroneous part of the nascent polynucleotide which result in the transition to the active state of RNAP.}
\label{sahoo}
\end{figure}

Very recently the effects of RNAP traffic congestion on the backtracking 
of individual RNAPs and kinetic proofreading have also been studied 
theoretically \cite{sahoo11}. In the Sahoo-Klumpp model of nucleolytic 
proofreading during transcription \cite{sahoo11}, a rigid rod that 
represents a RNAP and has step size unity in the units of a single base. 
Upon incorporation of an incorrect nucleotide, it makes a transition to 
the ``error state'' with rate $p$ without forward translocation 
(see fig.\ref{sahoo}). 
Alternatively, even after such a misincorporation, it can translocate 
forward with a rate that is lower than its normal rate of forward 
translocation (which is accompanied by the incorporation of a correct 
nucleotide). Once in the ``error state'', the RNAP can backtrack; in 
this state it moves in a diffusive manner. Elongation can resume along 
two alternative routes. The RNAP can either regain its active position 
and then resume elongation (thereby leaving the erroneoualy incorporated 
nucleotide intact) or cleave the transcript at its backtracked position 
and occupy the newly created active state (thereby resuming elongation 
after correcting the error) (see fig.\ref{sahoo}). 
Sahoo and Klumpp \cite{sahoo11} calculated 
the fraction of the errors that are corrected by backtracking and 
transcript cleavage. This process, when carried out by a single RNAP 
without hindrance or obstruction from any other RNAP, is similar to 
that studied earlier by Voliotis et al.\cite{voliotis09} by a slightly 
different set of mathematical steps. However, Sahoo and Klumpp studied 
the more general scenario by taking into account the effects of 
steric interactions of the RNAPs in a congested RNAP traffic. In such a 
situation success of error correction hinges on the rate of cleavage; this 
rate has to be sufficiently high so that erroneously incorporated 
nucleotide is cleaved before the backtracked RNAP get reactivated by a 
trailing RNAP.

\section{DDDP and DNA replication} 

Most of the DNA polymerases have a ``cupped right hand''-like structure 
where its sub domains can be identified as palm, thumb and finger domain. 
Shapes and sizes of these sub domains vary extensively from polymerase 
to polymerase but overall structure remains the same. The template strand 
enters through the finger domain and exits from the thumb domain, while 
the dNTP binds between finger and palm domain.

\subsection{Single DDDP: speed and fidelity of DNA replication} 

In their pioneering {\it in-vitro} experiments on DDDP, Wuite et al. 
\cite{wuite00} applied tension on a ssDNA molecule, that served as 
the template, by holding it with a micro pipette at one end and an 
optical trap on the other. Similar experiments were carried our, 
almost simultaneouly, by Maier et al.\cite{maier00} 
on a different DNAP and using a magnetic trap. 
In these experiments, the ssDNA was converted into a dsDNA by the 
DDDP. Interestingly, the average rate of replication $k(F)$ was 
found to vary nonmonotonically with the tention $F$ \cite{wuite00,maier00}.
The observed trend of variation of the replication rate was explained 
in terms of the differences in the force-extension curves of ssDNA 
and dsDNA \cite{wuite00,maier00}.  
An alternative model developed by Goel et al.\cite{goel01,goel03} 
has been further elaborated by an atomistic model, that was studied 
computationally, by Andricioaei et al.  \cite{andricioaei04}.

A DDDP is a dual-purpose enzyme that plays two opposite roles in two 
different circumstances during DNA replication. It plays its normal 
role as a polymerase catalyzing the {\it elongation} of a new DNA 
molecule. However, it can switch its role to that of a exonuclease 
catalyzing the {\it shortening} of the nascent DNA by cleavage of 
the nucleotide at the growing tip of the elongating DNA \cite{krantz10}.
The two distinct sites on the DNAP where, respectively, polymerization
and cleavage are catalyzed, are separated by 3-4 nm
\cite{ibarra09}. 
The nascent DNA is transferred back to the site of polymerization after 
cleaving the nucleotide from its growing tip. The elongation and cleavage
reactions are thus {\it coupled} by the transfer of the DNA between
the sites of polymerase and exonuclease activity of the DNAP. However,
the physical mechanism of this transfer is not well understood
\cite{xie09}.
``Exo-deficient'' mutants and ``transfer-deficient'' mutants have been 
used to understand the interplay of exonuclease and transfer processes 
on the platform of a single DDDP \cite{ibarra09}. 

Normally, transfer of the nascent DNA from the polymerase site to 
the exonuclease site takes place upon incorporation of a wrong 
nucleotide so that the misincorporated nucleotide can be cleaved off. 
However, in spite of this quality control system, some misincorporated 
nucleotides can escape cleavage; such {\it replication error} in the 
final product is usually about $1$ in $10^{9}$ nucleotides. Moreover, 
one cannot rule out the possibility of a similar transfer, albeit 
rarely, even after the incorporation of a correct nucleotide. If 
in this process the correct nucleotide is erroneously cleaved off 
before getting transferred back to the polymerase site, such ``futile'' 
cycles of the DDDP would unnecessarily slow down the replication 
process \cite{fersht82}.

\begin{figure}[ht]
\begin{center}
\includegraphics[angle=0,width=0.6\columnwidth]{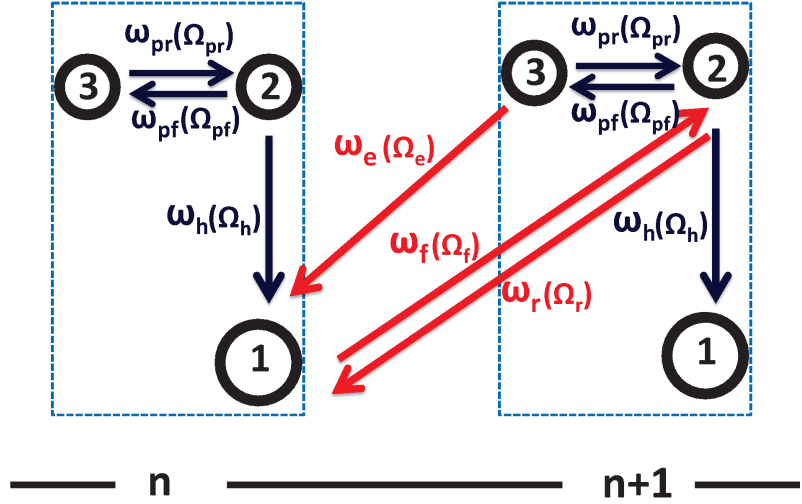}
\end{center}
\caption{Full mechano-chemical cycle in the model of DNAP developed by 
Sharma and Chowdhury \cite{sharma12} (see the text for details).}
\label{dnap1}
\end{figure}

Fig.\ref{dnap1} depicts, a minimal three-state kinetic model \cite{sharma12} 
of DNA polymerase on leading strand, where the continuous replication take 
place. In this model correct and incorrect nucleotides follow the same 
kinetic path but the rate constants for chemical transition differs 
significantly, making the correct incorporation more favorable. Rate 
constants for correct and incorrect nucleotides are represented by $\omega$ 
and $\Omega$, respectively, and the same subscripts are used for the same 
type of transitions. In the figure \ref{dnap1}, chemical state 1 represents 
the state where DNA polymerase is ready for the next round of elongation 
cycle. The transition $1 \rightarrow 2$ stands for polymerization step, 
occurs with rate $\omega_{f}(\Omega_{f})$ for correct(incorrect) nucleotides. 

\begin{figure}[ht]
\begin{center}
\includegraphics[angle=0,width=0.9\columnwidth]{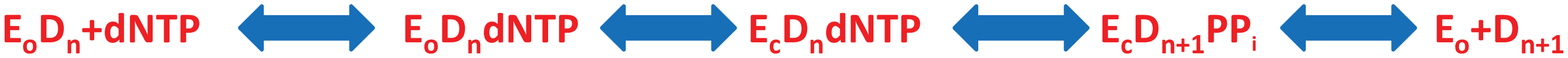}
\end{center}
\caption{Sub-steps in the transition $1\rightarrow 2$ shown in 
fig.\ref{dnap1} (see the text for details).}
\label{dnap2}
\end{figure}

In principle, the transition $1\rightarrow 2$ can be divided into more than 
one sub steps\cite{Johnson10}. These sub steps are shown in figure \ref{dnap2}. 
In figure \ref{dnap2}, $E_{c}$ and $E_{o}$ denote the closed and open 
conformation of the DNA polymerase, whereas $D_{n}$ represent the position 
of the catalytic site. When a substrate (dNTP) binds with DNAP, its binding 
energy transforms the DNAP from an open configuration to a close configuration. 
This new configuration of the DNAP favors the formation of diester bond. 
As a result of the polymerization reaction the nascent DNA is elongated by 
one nucleotide. This sub-step is followed by switching of the DNAP to the 
open conformation and the release of $PP_{i}$, thereby completing the 
$1\rightarrow 2$ transition. Random fluctuations between open and closed  
conformations of DNAP may drive the reaction in backward direction causing 
the disassociation of dNTP by rate $\omega_{r}$($\Omega_{r}$) for correct 
(incorrect) nucleotide. 

The transition $2$($E_{o}+D_{n+1}$) $\rightarrow$ $1(E_{o}D_{n+1})$ 
symbolizes the relaxation of the last incorporated nucleotide with rate 
$\omega_{h}(\Omega_{h})$ for correct (incorrect) nucleotide incorporated. 
Chemical state $2(E_{o}+D_{n+1})$ may activate the exonuclease mode of the 
enzyme (state 3) by transferring the growing DNA chain to exonuclease site 
at the rate $\omega_{pf}$($\Omega_{pf}$) for correct(incorrect) nucleotide. 
The active exonuclease site can cleave the last incorporated nucleotide 
with rate $\omega_{e}$($\Omega_{e}$) for correct (incorrect) incorporation, 
alternatively it may also switch back to polymerase active mode with rate 
$\omega_{pr}$($\Omega_{pr}$) for correct (incorrect) incorporation.  
For this model, Sharma and Chowdhury \cite{sharma12} derived the 
distribution of the dwell times and that of the exonuclease turnover 
times. The exact analytical expressions for these distributions display 
the effect of the coupling of two different enzymatic activities of a 
DNAP, namely, its polymerase activity and exonuclease activity.


\begin{figure}[ht]
\includegraphics[angle=0,width=0.8\columnwidth]{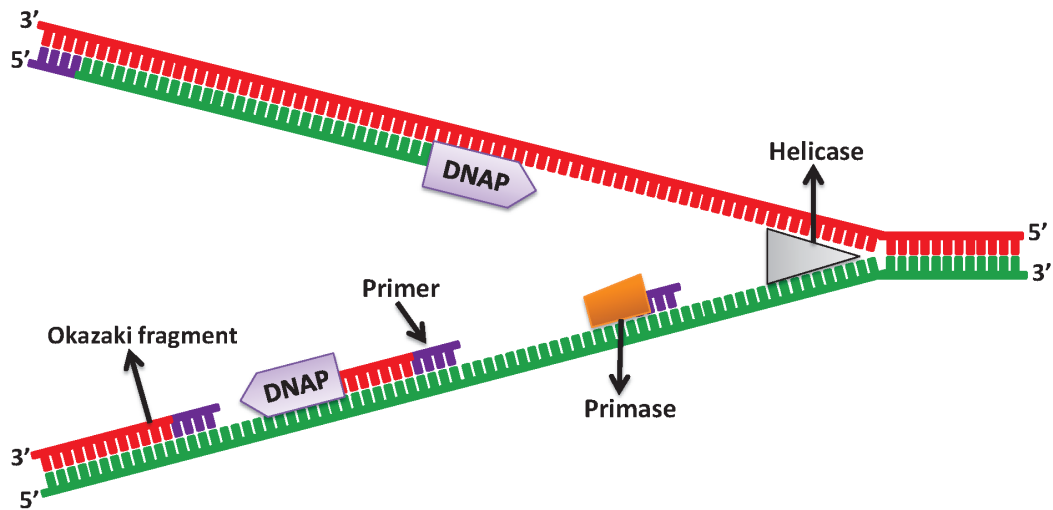}
\caption{DNA replication process (see text for details).}
\label{okazaki}
\end{figure}

Unlike RNAP, the DNAP is not capable of helicase activity. Therefore, 
ahead of the DNAP, a helicase progressively unzips the dsDNA thereby 
exposing the two single strands of DNA which serve as the templates 
for DNA replication (see fig.\ref{okazaki}). For the processive 
translocation of a DNAP on its template, it needs to be clamped with 
a ring-like ``DNA clamp'' 
\cite{bloom01}. 
The assembly of a DNA clamp is assisted, in turn, by a ``clamp loader'' 
in a ATP-dependent manner 
\cite{bloom06,jeruzalmi02,indiani06,ellison01,barsky05}.   
DNAP cannot initiate replication on its own and requires priming by 
another enzyme called primase (see fig.\ref{okazaki})
\cite{frick01}.
Thus, DDDPs alone cannot replicate the genome; together with DNA clamp 
and clamp loader, DNA helicase and primase, it forms a large multi-component  
complex machinery which is often referred to as the {\it replisome} 
\cite{baker98,benkovic01,johnson05,davey00,frouin03,herendeen96,thommes90,bell02,garg05,barry06,mcgeoch08,oijen10,perumal10,patel11}. 
The spatio-temporal coordination of the operation of the different 
components of the replisome during DNA replication is the most 
interesting aspect of its operational mechanism.

\subsection{Coordination of two replisomes at a single fork} 

DNA replication is more complex than transcription. 
Two DDDPs have to replicate these two complementary strands of DNA. 
However, each DDDP is capable of translocating only unidirectionally 
($5' \to 3'$) for elongating the corresponding product strand. As a 
result, one of the strands (called the ``leading strand'') is synthesized 
processively, whereas the ``lagging strand'' is replicated discontinuously; 
processing of these ``Okazaki fragments'' into a continuous DNA strand 
takes place in three steps catalyzed by three enzymes which are 
not part of the replisome (see fig.\ref{okazaki}).

How are the replication of the leading and lagging 
strands maintain tight coordination as the replication fork moves 
forward? Does the DNAP on the lagging strand polymerize at a faster 
rate than that on the leading strand so as to make up for the time 
lost in the priming and in re-starting DNA elongation thereby 
enabling it to catch up the DNAP on the leading strand? Or, does the 
DNAP on the leading strand make a pause at the replication fork during 
the interval between the end of synthesis of one Okazaki segment and 
the beginning of that of the next Okazaki segment on the lagging 
strand?  Experimental investigations, particularly, single-molecule 
experiments, have started addressing these questions in the recent 
years. For example, it has been found that the primase acts, at 
least effectively, as a molecular ``brake'' preventing the 
leading-strand synthesis from outpacing the lagging-strand synthesis 
of DNA \cite{lee06,lee10}.

\section{RDRP of RNA viruses and RNA replication} 

RNA viruses contain a small genome that is usually not longer than 
30kb. It may consist of either a ssRNA or a dsRNA. 
In some viruses RNA genome consists of more than one segment 
whereas in others it is a single segment. In multi segment RNA viruses, 
all the genome segments are replicated independently.

In spite of strong resemblance of the overall shape of all the RDRPs
with a ``cupped right hand'', viral RDRPs have some special architectural
features \cite{blumenthal79,ortin06,ng08,roy08}. 
The most notable distinct feature of these polymerases is that, in 
contrast to the ``open hand'' shape of the other polynucleotide 
polymerases, the RDRP resembles a ``closed hand''. The closing of the 
``hand'' is achieved by loops, called ``fingertips'', which protrude 
from the fingers and connect with the thumb domain at their other end. 
The fingertip region forms the entrance of the channel where the RDRP 
binds with the RNA template. In addition, there is a small positively 
charged tunnel through which the nucleotide monomers, required for 
elongation of the RNA, enter.

RDRP reads the template by moving along the $3’$ to $5’$ 
direction on the template RNA while simultaneously synthesizing the 
complementary RNA from $5’$ to $3’$ direction\cite{orta06}.
The genome of some of the viruses consist of double stranded RNA;
the corresponding RDRP have some additional unique structural elements
which unzip the two strands and feed the appropriate strand to the
catalytic site.

RDRPs in different systems are known to adopt distinct strategies of 
initiation \cite{alberdina04}; these include both 
(i) {\it Primer independent initiation} as well as 
(ii) {\it Primer dependent initiation}. 
In the latter case several different potential sources of primers have 
been identified. For example, 
(a) an oligonucleotide discarded as a result of abortive initiation 
can serve as a primer;  
(b) an oligonucleotide generated by the cleavage of the 5’ end of a 
mRNA of the host cell can also be exploited as a primer; 
(c) the 3' end of the looped template itself can be utilized as the 
primer.

\section{RDDP of retroviruses and reverse transcription}

\begin{figure}[ht]
\includegraphics[angle=0,width=0.6\columnwidth]{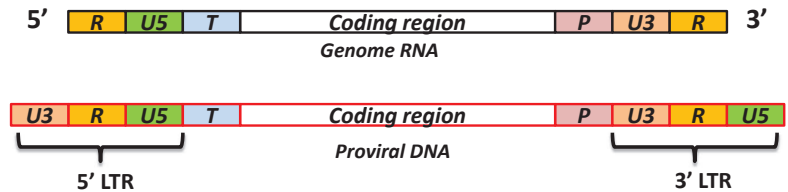}
\caption{Sequence of the genomic RNA and proviral DNA are shown 
schematically (adapted from \cite{dimmock07}). Different regions of 
these sequences are represented by different colors. R is the directly 
repeated sequence found at both termini of the genome RNA but is 
internal to the proviral DNA. U3 and U5 are unique sequences located 
near the 3' and 5' ends of the genomic RNA, respectively. Combination 
of U3, R and U5 is known as long terminal repeats (LTR). T is 
complimentary to the host tRNA (primer for reverse transcription) and 
P is polypurine rich region, which is relatively resistant to the RNaseH 
activity of RT.
}
\label{rt1}
\end{figure}

Recall that transcription is the process whereby a RNA strand is 
polymerized using a DNA strand as the template. Therefore, the 
reverse process, i.e., polymerization of a DNA strand using RNA 
template was given the name ``reverse transcription''  
and the corresponding polymerase is called a reverse transcriptase 
(RT).

Reverse transcription is a crucial step in the life cycle of 
retroviruses \cite{galetto09}. 
Research on retroviruses got an enormous boost in the mid-1980s after 
AIDS became one of the main focus of research in virology and medicine. 
Obviously, most of the research on RT over the last four decades has been 
dominated by studies of the RT of human immunodeficiency virus (HIV) 
\cite{wendeler09}. 
RT of HIV is one of key targets for the some of the drugs which are 
being tried against AIDS.
Neverthless, impressive progress have been made also in understanding 
the structure and function of non-HIV RTs \cite{hizi08}.

\begin{figure}[ht]
\includegraphics[angle=0,width=0.45\columnwidth]{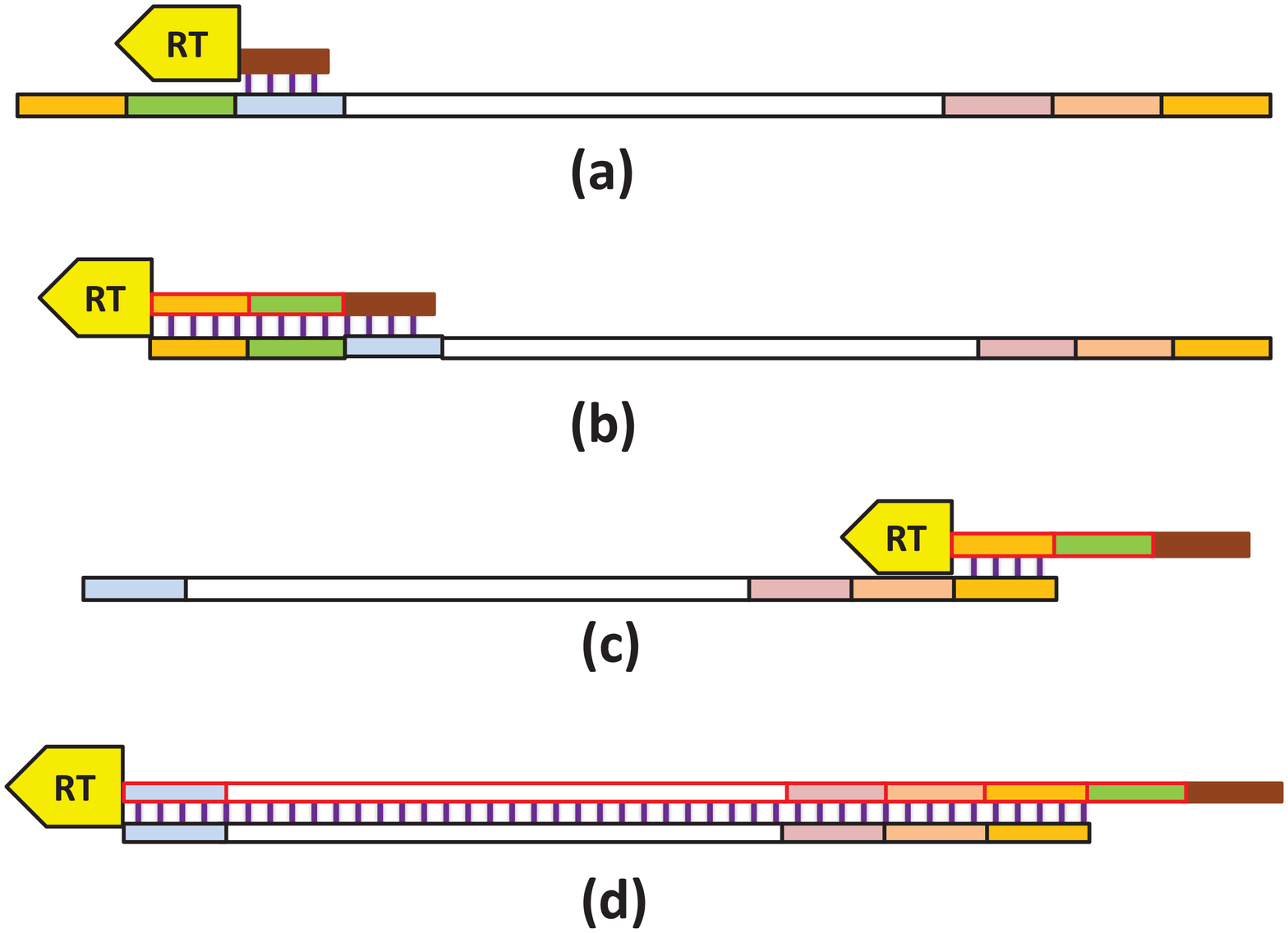}
\includegraphics[angle=0,width=0.45\columnwidth]{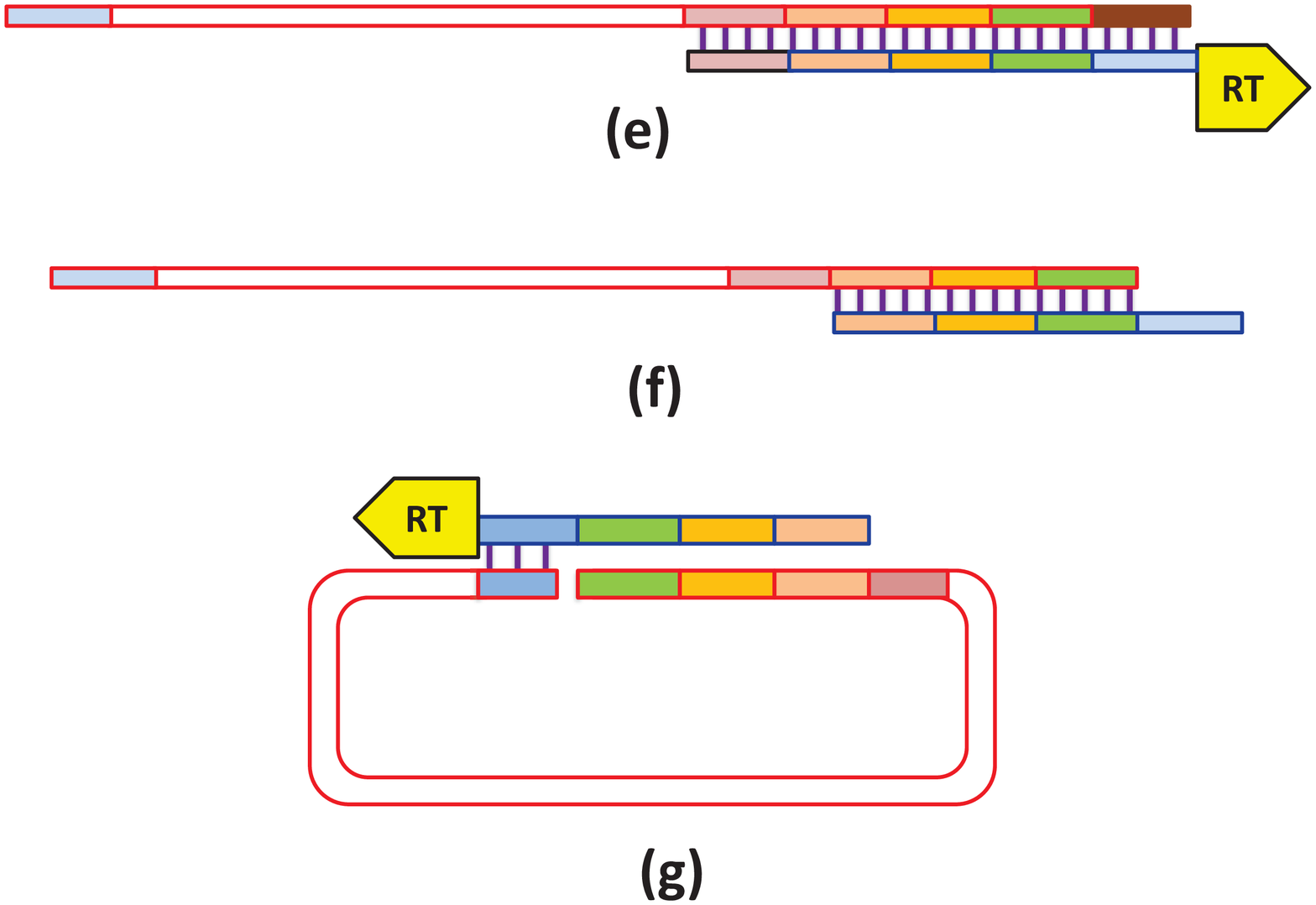}
\includegraphics[angle=0,width=0.45\columnwidth]{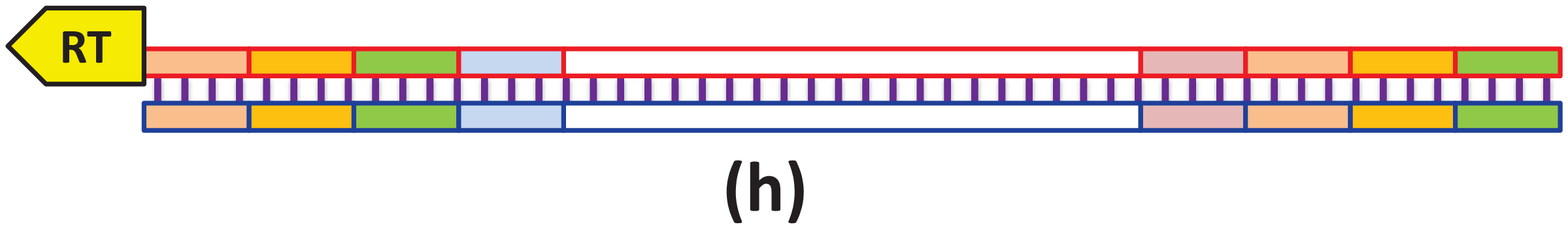}
\caption{Schematic diagram of reverse transcription (adapted from \cite{beilhartz10}) Different polynucleotids are represented by the boundaries of different colors.
RNA genome have black boundary whereas positive sense and negative sense DNA
strands are covered by red and blue boundaries, respectively.
(a) tRNA molecule forms the base pairs with T and works as a
primer to initiate the reverse transcription.
(b) Primer is extended by RT till the 5' end of the genome.
(c) While elongating the DNA, RNase H activity of RT digests the template RNA by leaving the DNA free from base pairing. So the DNA re-associate with the sequence R at the 3' end of the RNA genome, known as first template switch.
(d) RT elongates the DNA till 5' end of the genome.
(e) RNA exist in the DNA-RNA duplex is degraded by the RNaseh activity of the RT
except the sequence P which works as a primer for the initiation of the positive strand synthesis. RT elongates the DNA  till the end of template at tRNA.
(f) Now the tRNA and region P are degraded by RNase H activity of RT.
(g) The DNA strand form a loop which gives the opportunity to form a base pair between
the T region of both strands.
(h) After the positive strand transfer both DNA strands are elongated till the ends of their template.
}
\label{rt2}
\end{figure}

Strictly speaking, a RT performs three distinct tasks \cite{herschhorn10} 
(see fig.\ref{rt2}): 
(i) reverse transcription: RT polymerizes a DNA strand using the genomic 
RNA as the corresponding template (i.e., the process from which it derives 
its name); 
(ii) RNaseH activity: RT cleaves the RNA strand of the DNA-RNA duplex 
formed by the process (i) above; 
(iii) DNA polymerase: RT also plays the role of a DDDP by catalyzing the 
polymerization of a DNA strand that is complementary to the DNA strand 
synthesized in the process (i) above.\\ 
Thus, a RNA strand, which constitutes the viral genome, serves only as 
the initial template and the final product of reverse transcription is 
called a proviral DNA (see fig.\ref{rt1}).  

Structural studies of RT have revealed that DNA polymerase domains of 
the RT are similar to the DNA polymerases of living cells. But, the RT 
also has some additional distinct features which are responsible for 
its RNaseH activity. The Catalytic domains that perform the Polymerase 
and RNaseHase activities of a RT are spatially distinct, but are linked 
through the ``connection subdomain'' \cite{herschhorn10}.
RT of HIV binds with substrate in two different orientations each of 
which is capable of DNA polymerase and RNaseH activities. Switching 
of the orienations switch the RT activity \cite{Abbondanzieri08}. 
This switching ocuurs spontaneously and it is regulated by the small 
ligand molecules.

Majority of the RTs use host tRNA as the required primer. But, some RTs 
use other resources for priming \cite{levin97}. 
RT is a deficient in proofreading capabilities \cite{mansky98}. 
Consequently, the rate of reverse-transcriptional error is quite high. 
Since completion of the integration of the provirus into the host genome 
involves several steps driven by the RT, the errors of each step add up. 
Higher mutation rate in retroviruses has severe consequences- the virus 
uses it as one of its survival strategies against the host defence system 
whereas the host immune systems finds it difficult to recognize it. 
A kinetic model for the stochastic description of the process of 
reverse transcription by RT has been formulated \cite{sharma12b}

\section{Interference of transcription and replication: traffic rules for RNAP-DNAP collision} 

\begin{figure}[ht]
\includegraphics[angle=0,width=0.8\columnwidth]{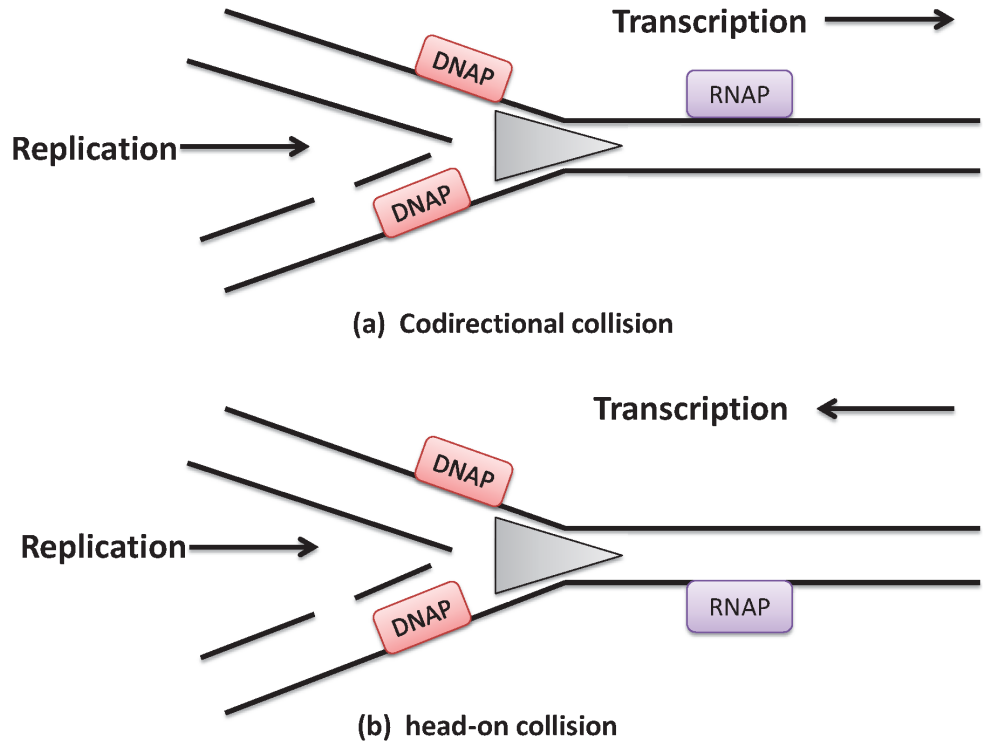}
\caption{Schematic representation of collision between replication fork and TEC (a) codirectional collision: replication fork and TEC move in the same direction and RNAP transcribes the leading strand. (b) head-on collision: replication fork and TEC move in opposite direction and RNAP transcribes the lagging strand (inspired by ref.\cite{soultanas11}).}
\label{collision}
\end{figure}

Transcription of a gene is carried out a large of times during the life
time of a single cell. In contrast, a distinct feature of DNA replication
is that, during its lifetime, a cell must not replicate its genome more
than once. Only recent investigations have explored how cell achieves
this requirement.
Replication of bacterial genome (e.g., E-coli) proceeds {\it bidirectionally} 
from the initiation site (oriC). Each replication fork, alongwith its 
replisome synthesizing the leading and lagging strands approaches the same 
termination site (terC), but from opposite directions. As the replication 
forks move, there is a possibility for collision with a TEC on the way if 
a gene is getting transcribed simultaneously (see Fig.\ref{collision}).

When a replication fork follows a leading TEC, the DNAP remains stalled 
as long as the leading RNAP pauses on its track; however, the DNAP resumes 
replication once the RNAP starts moving forward again  \cite{arnanz97}. 
Similarly, if the replication fork collides with the TEC {\it head-on} 
from opposite direction, the DNAP can bypass the moving RNAP \cite{arnanz99}.

\section{Ribosomes and polymerization of polypeptides} 

Ribosome, one of the largest and most sophisticated macromlecular 
machines within the cell, polymerizes polypeptides using a mRNA 
as the corresponding template 
\cite{spirin99,spirin02,spirin04,spirin09,sanbonmatsu12}. 
In each successfully completed mechano-chemical cycle of a ribosome 
two molecules of guanosine triphosphate (GTP) are hydrolyzed into 
 guanosine diphosphate (GDP). Moreover, one of the steps of this 
cycle needs the assistance of specifically prepared molecular 
assembly whose preparation also involves hydrolysis of a molecule 
of ATP. Because of these energy-consuming steps involved in the 
operation of a ribosome, it is regarded as a molecular motor   
\cite{abel96}.

A ``slippage'' of the reading frame by $3n$ nucleotides, where $n$ is 
an integer, would result in missing $n$ amino acids without affecting 
the identity of the other amino acids. However, if the slippage is 
not a multiple of $3$ nucleotides, the entire sequence of amino 
acids after the slippage would be different from the coded sequence.

Aminoacyl tRNA synthetase (aa-tRNAsynth) ``charges'' a tRNA molecule 
with a amino acid. In order to ensure high fidelity of translation, 
the aa-tRNAsynth must have high specificity for its two substrates, 
namely, tRNA and amino acid \cite{ling09}. The error committed by 
aa-tRNAsynth never exceeds once in $10^4$ enzymatic cycles.
Interestingly, aminoacyl-tRNA synthetase and DNAP share some common 
mechanisms to ensure translational and replicational fidelities, 
respectivly \cite{francklyn08}.

It has been argued \cite{spirin99} that the energy of the chemical 
bond between the amino acid and tRNA is later used by the ribosome for 
forming a peptide bond between this amino acid and the nascent polypeptide. 
The free energy released by the hydrolysis of two GTP molecules are utilized, 
respectively, in selecting the correct aa-tRNA and the release of the 
deacylated tRNA into the surrounding aqueous medium.

\subsection{Composition and structure of a single ribosome}

Even in the simplest organisms like single-cell bacteria, a ribosome is 
composed of few rRNA molecules as well as several varieties of protein 
molecules. 
The structure of both bacterial and eukaryotic ribosomes have been 
revealed by extensive detailed investigation over several years by 
a combination of X-ray diffraction, cryo-electron microscopy, etc. 
\cite{westhof00,moore03,steitz08,steitz10,yonath02,yonath10,ramakrishnan02,schmering09,ramakrishnan10,frank00,frank09,frank10,yusupov01,blaha04}. 
For this achievement, V. Ramakrishnan, T.A. Steitz and A. Yonath 
shared the Nobel prize in chemistry in 2009 
\cite{ramakrishnan10,steitz10,yonath10}. 
For many years the mechano-chemical kinetics of ribosomes have been 
investigated by studying bulk samples with biochemical analysis as 
well as the structural probes mentioned above. Only in the last few  
years, it has been possible to observe translation by single isolated 
ribosome {\it in-vitro} 
\cite{marshall08,blanchard04,uemura07,munro08,munro09,blanchard09,uemura10,aitken10,vanzi07,wang07c,wen08,tinoco09}.

Ribosomes found in nature can be broadly divided into two classes: 
(i) prokaryotic $70$S ribosomes, and (ii) eukaryotic $80$S ribosomes; 
the numbers $70$ and $80$ refer to their sedimentation rates in the 
Svedberg (S) units. In the earliest electron microscopy the prokaryotic 
and eukaryotic ribosomes appeared to be approximately spherical 
particles of typical diameters in the ranges $20-25$ nm and $20-30$ nm, 
respectively. In these electron micrographs a visible groove was 
found to divide each ribosome into two unequal parts; the larger and 
the smaller parts are called, for obvious reasons, large and small 
subunit, respectively. The sizes of the large and small subunits of 
the $70$S ribosome are $50$S and $30$S respectively, whereas those of 
the $80$S ribosome are $60$S and $40$S, respectively.

The small subunit binds with the mRNA track and assists in decoding the 
genetic message encoded by the codons (triplets of nucleotides) on the mRNA. 
The ``head'' and the ``body'' are the two major parts of the {\it small} 
subunit. Two major lobes, which sprout upward from the ``body'', are 
called the ``platform'' and the ``shoulder'', respectively. The decoding 
center of the ribosome lies in the cleft between the ``platform'' and 
the ``head'' of the small subunit. The incoming template mRNA utilizes a 
``channel'' formed between the ``head'' and the ``shoulder'' as a conduit 
for its entry into the ribosome. Through the cleft between the ``head'' 
and the ``platform'' the mRNA exits the ribosome.

But, the actual polymerization of the polypeptide takes place in the large 
subunit. The characteristic ``crown-like'' architecture of the {\it large} 
subunit arises from three protuberances. On the flat side of the large 
subunit exists a ``canyon'' that runs across the width of the subunit 
and is bordered by a ``ridge''. Halfway across this ridge, a hole leads 
into a ``tunnel'' from the bottom of the ``canyon''. This ``tunnel'' 
penetrates the large subunit and opens into the solvent on the other 
side of the large subunit. This ``tunnel'' serves as the conduit for 
the exit of the nascent polypeptide chain. This ``tunnel'' is 
approximately 10 nm long and its average width is about 1.5 nm.

Several intersubunit ``bridges'' connect the two subunits of each
ribosome. This bridges are sufficiently flexible so that relative
movements of the two subunits can take place in each cycle of the
ribosome. The intersubunit space is large enough to accomodate just 
three tRNA molecules which can bind, at a time, with the three 
binding sites designated as E, P and A. Moreover, the shape of the 
intersubunit space is such that it allows easy passage of the 
L-shaped tRNA molecules. The operations of the two subunits are 
coordinated by the tRNA molecules.

\subsection{Mechano-chemical kinetics of a single ribosome: speed and fidelity of translation}

A ribosome may translate at the rate of a few codons to few tens 
of codons per second.
Just like the synthesis of polynucleotides (e.g., transcription and 
replication), synthesis of polypeptides (i.e., translation) also 
goes through three stages, namely, {\it initiation}, {\it elongation}, 
and {\it termination}.

\begin{figure}[ht]
\begin{center}
\includegraphics[angle=-90,width=0.4\columnwidth]{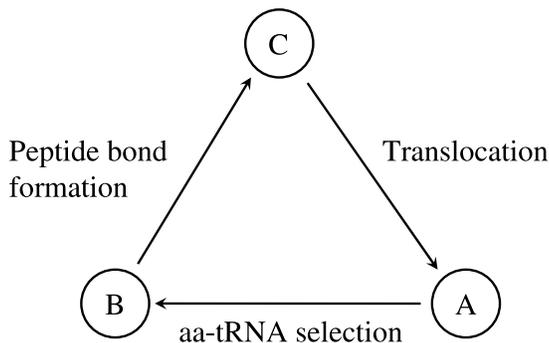}
\end{center}
\caption{Pictoral depiction of the three main stages in the
chemo-mechanical cycle of a single ribosome (see the text for
details).
}
\label{fig-3state}
\end{figure}

During the elongation stage, while translating a codon on the mRNA 
template, the three major steps in the mechano-chemical cycle of a 
ribosome are as follows (see fig.\ref{fig-3state}): 
In the first, based on matching the codon with the anticodon on the 
incoming aa-tRNA, the ribosome {\it selects} the correct amino acid 
monomer that, according to the genetic code, corresponds to this  
codon. Next, it catalyzes the chemical reaction responsible for the 
formation of the peptide bond between the nascent polypeptide and 
the newly recruited amino acid resulting in the {\it elongation} of 
the polypeptide. Final step of the cycle is {\it translocation} at 
the end of which the ribosome finds itself at the next codon and is 
ready to begin the next cycle.

Thus, each ribosome has three distinct functions which it performs on
each run along the mRNA track:\\
(i) it is a {\it decoding device} in the sense that it ``reads'' the
genetic message encoded in the sequence of codons on the template 
mRNA and selects the correct corresponding amino acid monomer;\\  
(ii) it is a ``{\it polymerase}'' because it elongates the polypeptide 
chain by one amino acid by catalyzing the formation of the peptide bond 
between the nascent polypeptide and the newly recruited amino acid;\\ 
(iii) it is a {\it motor} that steps forward by one codon on a mRNA 
strand by transducing chemical energy into mechanical work.\\
Interestingly, function (i) is performed exclusively by the smaller
subunit while the function (ii) is carried out in the larger subunit.
But, the function (iii) requires coordinated movement of the two 
subunits by tRNA molecules.

Elongation factors (EF), which are themselves proteins, play important 
roles in the control of the three major steps shown in fig.\ref{fig-3state}. 
During the process of checking its identity through the condon-anticodon 
matching (which takes place in the smaller subunit), the formation of 
the peptide bond is prevented by an elongation factor Tu (EF-Tu). 
However, once a cognate tRNA is identified, the smaller subunit sends 
a ``green signal'' (by a molecular mechanism that remains unclear), the 
EF-Tu separates out by a process driven by GTP hydrolysis thereby 
clearing the way for the peptide bond formation. Similarly, elongation 
factor G (EF-G) coordinates the translocation of the mRNA by one codon 
and the simultaneous movement of the tRNA molecules from one binding site
to the next one. Thus, the hydrolysis of the first GTP molecule is 
exploited for the selection of the correct aminoacyl-tRNA at the A site 
whereas that of the second GTP molecule is utilized to release of the 
deacylated tRNA molecule from the E site.

\begin{figure}[ht]
\begin{center}
\includegraphics[angle=0,width=0.5\columnwidth]{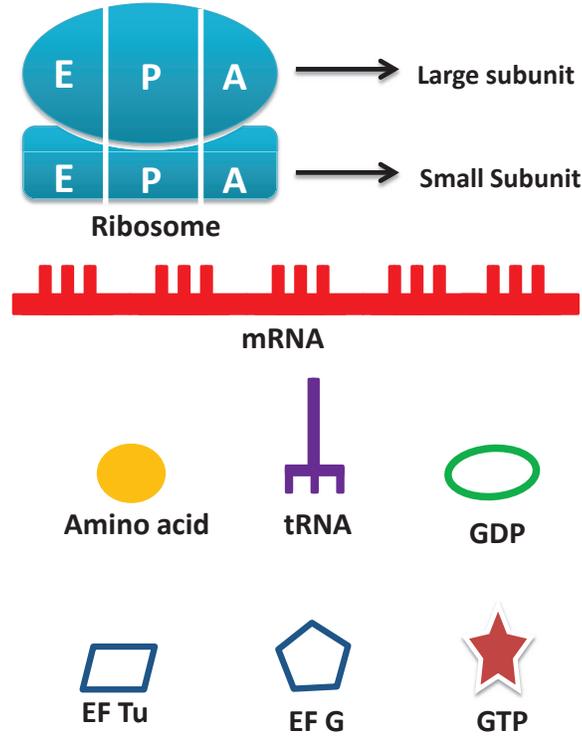}
\end{center}
\caption{Pictorial depiction of some of the key devices that participate in translation (see text for detailed explanation).}
\label{components}
\end{figure}

Clearly, the 3-state cycle sketched in fig.\ref{fig-3state} is an 
oversimplified description of the mechano-chemical kinetics of a 
ribosome during the elongation stage. 
We'll see in this section that at least two of the three steps in 
fig.\ref{fig-3state} consist of important sub-steps. Moreover, the 
aa-tRNA selected (erroneously) by the ribosome may not be the 
correct (cognate) tRNA. Rejection of such non-cognate and near-cognate 
tRNAs by the process of kinetic proofreading leads to an alternative 
branch completion of which ends up in a futile cycle. The roles played 
by some of the key devices sketched in fig.\ref{components}, which we 
explain below, have to be captured by a more detailed model of translation.
Next we review such a modeling strategy.

\subsubsection{Selection of amino-acid: two sub-steps}

\begin{figure}[ht]
\begin{center}
\includegraphics[angle=0,width=0.5\columnwidth]{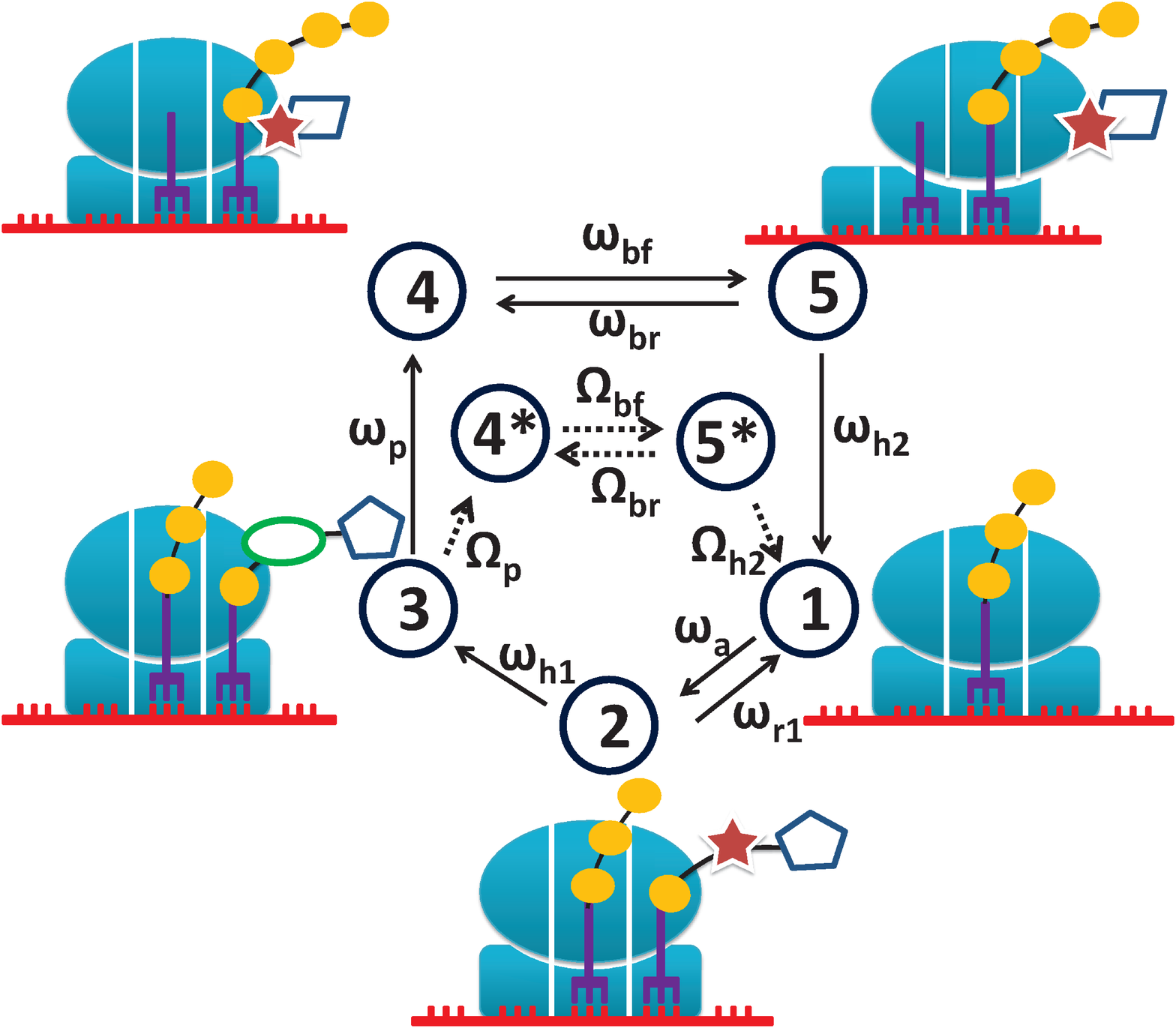}
\end{center}
\caption{Pictorial depiction of the full mechano-chemical cycle of the ribosome}
\label{ribo_model}
\end{figure}

Selection of an amino acid by a ribosome is a two-step process and these steps are (i) initial selection and (ii) kinetic proofreading. Elongation cycle of the ribosome starts with chemical state 1, shown in figure \ref{ribo_model}. In this state growing polypeptide chain is attached with the site P of the large subunit of ribosome whereas site E and A remain empty. An initial selection begins when a tRNA along with an amino acid subunit, elongation factor EF-Tu and a GTP molecule binds with the site A of the large subunit. This binding result in the transition to chemical state 2 with rate $\omega_{a}$.
All species of tRNAs, depending upon their relative concentration in the solution, compete with each other to bind with the ribosome, but in order to ensure the optimum fidelity, most of the non cognate and some near cognate tRNAs are rejected on the basis of condon-anticodon matching. This rejection result in the $2\rightarrow1$ transition, with rate $\omega_{r1}$. If a tRNA is not rejected then this binding stabilizes the ribosome complex  and transmit a signal to EF-Tu  to hydrolyze the GTP molecule. Then the GTP molecule is hydrolyzed and  corresponding irreversible  transition $2 \rightarrow 3$, occurs with rate $\omega_{h1}$.  This hydrolysis process is followed by some structural rearrangements in ribosome, which result in the release of the non cognate and near cognate tRNAs along with the EF-Tu and GDP molecule. This irreversible step $3 \rightarrow 1$  occurs by rate $\omega_{r2}$ and often referred as kinetic proofreading.

\subsubsection{Peptide bond formation: peptidyl transfer}

Sometimes an incorrect amino acid can escape from the two-stage quality control mechanism of ribosome, resulting an error in the final product. If the ribosome selects the correct amino acid it follows the
main ($3\rightarrow 4\leftrightarrow 5 \rightarrow 1$) pathway but if the selected amino acid is incorrect then the relatively slow, branched ($3\rightarrow 4^{*}\leftrightarrow 5^{*} \rightarrow 1$) pathway is followed.
In the next step amino acid is bonded with growing polypeptide chain, and the GDP molecule as well as the protein elongation factor EF-Tu are released. Next, a new protein elongation factor EF-G along with along with a GTP molecule binds with the ribosome. For the correct amino acid this transition ($3\rightarrow 4$) take place with rate $\omega_{p}$, whereas for an incorrect amino acid it ($3\rightarrow 4^{*}$) occurs with rate $\Omega_{p}$. Note that the only difference between 4(5) and $4^{*}(5^{*})$ is that last incorporated nucleotide is correct in 4(5) but incorrect in  $4^{*}(5^{*})$.

\subsubsection{Translocation: two sub-steps?}

The next transition $4 \leftrightarrow 5$ ($4^{*} \leftrightarrow 5^{*}$) is the back and forth spontaneous Brownian rotation of the two subunits of the ribosome from a classical configuration (E/E,P/P,A/A) to hybrid configuration (E/P,P/A,A). This transition is purely driven by random thermal fluctuations without any external power supply. Forward transition occurs with rate $\omega_{bf}$($\Omega_{bf}$), while the backward transition occurs with rate $\omega_{br}$($\Omega_{br}$) for correct (incorrect) pathways. Finally the hydrolysis of the GTP along with translocation step results in $5 \rightarrow 1(5^{*} \rightarrow 1)$ transition with rate $\omega_{h2}$($\Omega_{h2}$) for correct(incorrect) amino acid. This GTP hydrolysis completes one elongation cycle.\\

\subsubsection{Dwell time distribution}

As observed in single molecule experiments 
\cite{marshall08,blanchard04,uemura07,munro08,munro09,blanchard09,uemura10,aitken10,vanzi07,wang07c,wen08,tinoco09},
the stochastic stepping of a ribosome is characterized by an alternating 
sequence of pause and translocation. The sum of the durations of a pause 
and the following translocation is defined as the time of a dwell of the 
ribosome at the corresponding codon. The codon-to-codon fluctuation in 
the dwell time of a ribosome arises from two different sources:
(i) {\it intrinsic} fluctuations caused by the Brownian forces as well 
as the low of concentrations of the molecular species involved in the 
chemical reactions, and (ii) {\it extrinsic} fluctuations arising from 
the inhomogeneities of the sequence of nucleotides on the template mRNA 
\cite{buchan07}. Because of the sequence inhomogeneity of the mRNA 
templates used by Wen et al. \cite{wen08}, the dwell time distribution 
(DTD) measured in their single-molecule experiment reflects a combined 
effect of the intrinsic and extrinsic fluctuations on the dwell time. 

The probability density $f_{dwell}(t)$ of the dwell times of a ribosome,
measured in single-molecule experiments \cite{wen08}, does not fit a 
single exponential thereby indicating the existence of more than one 
rate-limiting steps in the mechano-chemical cycle of each ribosome. 
Best fit to the corresponding simulation data was achieved assuming 
five different rate-determining steps \cite{tinoco09}. 

We'll now sketch a theoretical framework \cite{garai09,sharma11a} 
which provides an exact analytical expression for $f_{dwell}(t)$ 
in terms of the rate constants for the individual transitions in 
the mechano-chemical kinetics of a single ribosome. This scheme also 
involves essentially five steps in each cycle during the elongation 
stage of translation. However, for the sake of simplicity of 
analytical derivation of the exact expression for $f_{dwell}(t)$, 
this theory assumed the template mRNA to have a {\it homogeneous} 
sequence (i.e., all the codons of which are identical). 

If we start our clock ($t=0$) when a ribosome is in chemical state 1 at 
site $i$, then the time taken by the ribosome to reach the chemical state 
1 of $(i+1)th$ site is defined as the dwell time of the ribosome. 
Let us assume that $P_{\mu}(i,t)$ is the probability of finding the 
ribosome in $\mu_{th}$ chemical state at site $i$, at time $t$. 
For a single ribosome time evolution of the probability $P_{\mu}(i,t)$ 
are governed by the corresponding master equations \cite{sharma11a}. 

For the calculation of the dwell time following initial conditions are 
imposed:
\begin{equation}
P_{1}(i,0)=1 ~{\rm and}~ P_{2}(i,0)= P_{3}(i,0)= P_{4}(i,0)= P_{5}(i,0)= P_{4}^{*}(i,0)= P_{5}^{*}(i,0)=P_{1}(i+1)=0
\end{equation}
The probability density of the dwell times $f_{dwell}(t)\Delta t$ 
can be obtained from 
\begin{equation}
f_{dwell}(t)=\dfrac{dP_{1}(i+1,t)}{dt}=\Omega_{h2}P_{5}^{*}(i,t)+\omega_{h2}P_{5}(i,t)
\label{ft}
\end{equation}
In the translation process many ribosomes move on the same $mRNA$ track 
and each of them synthesizes a separate copy of the same protein. 
Their steric interaction can be captured by appropriately modifying 
the master equations.

\begin{figure}[ht]
\begin{center}
\includegraphics[angle=-90,width=0.5\columnwidth]{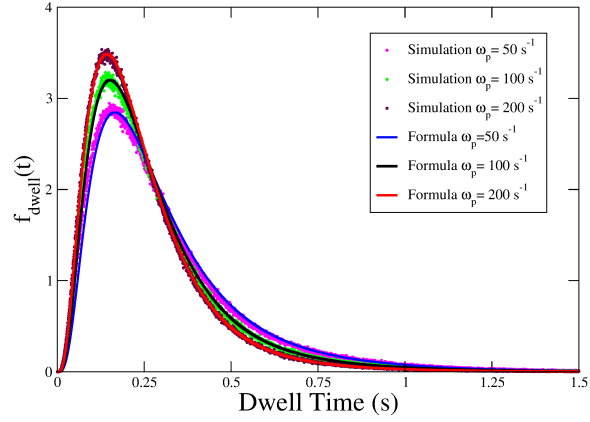}
\end{center}
\caption{Dwell time distribution of ribosome is plotted for a few different values of 
$\omega_{p}$. The values of other rate constants are $\omega_{a} = 25$s$^{-1}$,
$\omega_{r1} = 10$s$^{-1}$,
$\omega_{h1} = 25$s$^{-1}$,
$\omega_{r2} = 10$s$^{-1}$,
$\Omega_{p} = 40$s$^{-1}$,
$\omega_{bf} = \omega_{br} = 25$s$^{-1}$,
$\Omega_{bf} = \Omega_{br} = 10$s$^{-1}$,
$\omega_{h2} = 25$s$^{-1}$,
$\Omega_{h2} = 10$s$^{-1}$. The solid lines corresponds to the theoretical prediction whereas the discrete points corresponds to the data obtained by Monte Carlo simulation.}
\label{dt1}
\end{figure}
\begin{figure}[ht]
\begin{center}
\includegraphics[angle=-90,width=0.5\columnwidth]{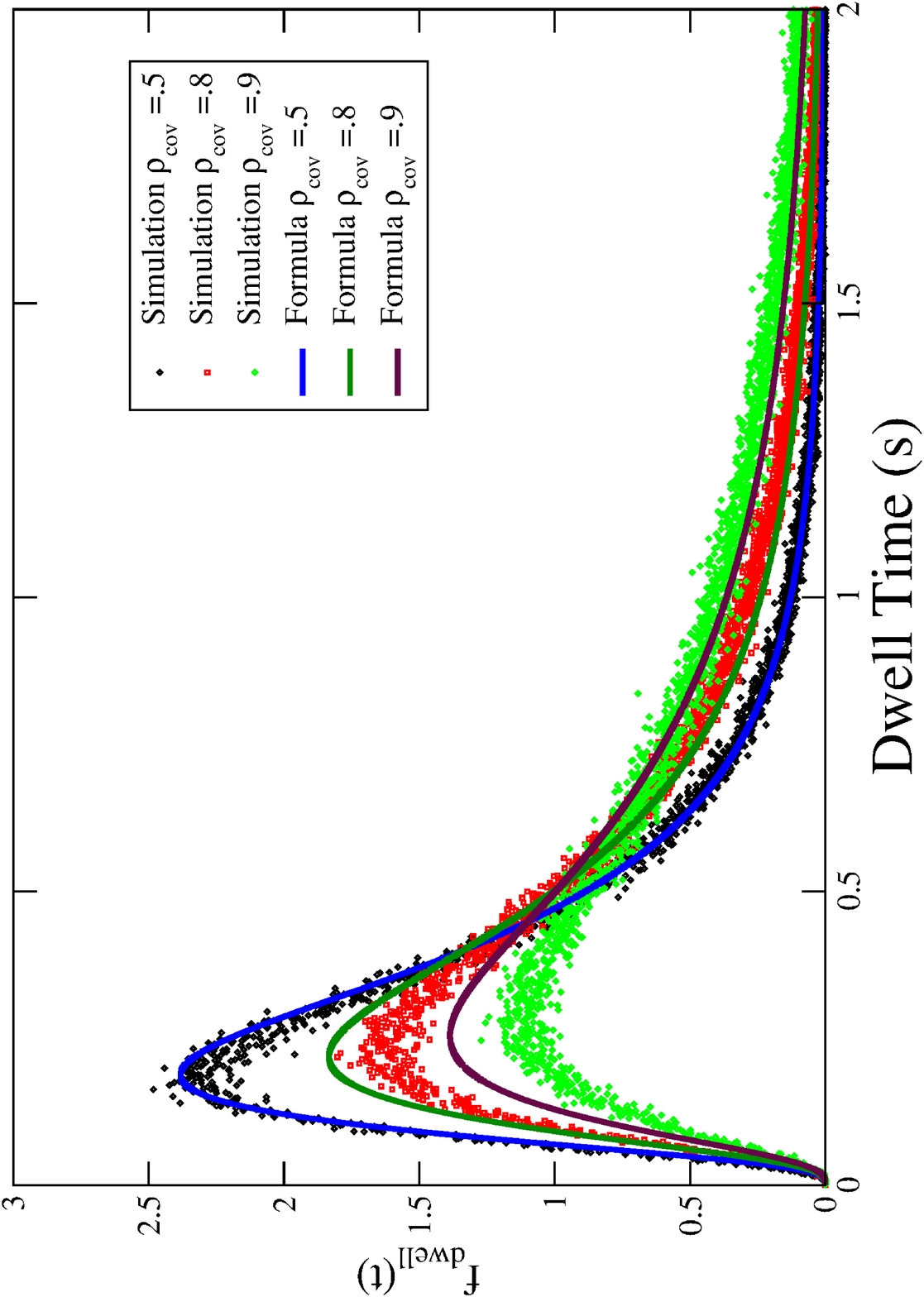}
\end{center}\caption{Dwell time distribution of ribosome is plotted for a few different values of 
$\rho_{cov}$. The values of other rate constants are $\omega_{a} = 25$s$^{-1}$,
$\omega_{r1} = 10$s$^{-1}$,
$\omega_{h1} = 25$s$^{-1}$,
$\omega_{r2} = 20$s$^{-1}$,
$\omega_{p} =50$s$^{-1}$
$\Omega_{p} = 40$s$^{-1}$,
$\omega_{bf} = \omega_{br} = 25$s$^{-1}$,
$\Omega_{bf} = \Omega_{br} = 10$s$^{-1}$,
$\omega_{h2} = 25$s$^{-1}$,$\Omega_{h2} = 10$s$^{-1}$. The solid lines corresponds to the theoretical prediction whereas the discrete points corresponds to the data obtained by Monte Carlo simulation
.}
\label{dt2}
\end{figure}

In figure \ref{dt1} dwell time distribution of a ribosome is plotted for 
a few different values of $\omega_{p}$.  The same (or closely related) 
set of values of the parameters were used earlier also for the purpose 
of plotting. The selection of these values were motivated by typical 
magnitudes reported in the literature for various steps of translation 
(most often in bulk measurements). Higher value of $\omega_{p}$ 
decreases the value of the most probable dwell time. In figure \ref{dt2} 
dwell time distribution of the ribosome is plotted for a few different 
values of $\rho_{cov}$. Due to the hindrance created by ribosome's mutual 
exclusion, higher value of $\rho_{cov}$ leads to a longer most probable 
dwell time. The deviation of the theoretical prediction from the Monte 
Carlo data at higher coverage density is a consequence of the mean field 
approximation which ignores the correlations.

\begin{figure}[ht]
\begin{center}
\includegraphics[angle=-90,width=0.6\columnwidth]{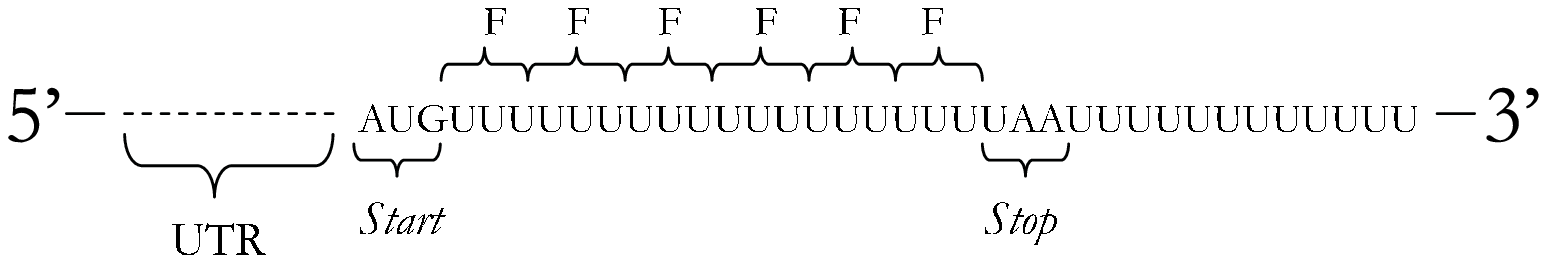}
\end{center}
\caption{A schematic description of a mRNA with homogeneous (poly-U)
coding sequence (adapted from refs.\cite{uemura10,aitken10}).
}
\label{fig-expt}
\end{figure}

The expression for $f_{dwell}(t)$ thus derived incorporates the 
effects of fluctuations that are strictly {\it intrinsic}. 
This model \cite{sharma11a} envisages a scenario that is very similar 
to the protocol used in some single-ribosome experiments \cite{uemura10} 
and is shown schematically in fig.\ref{fig-expt}. 

The coding sequence to be translated consists of $n_c$ number of 
identical codons. In the example shown of fig.\ref{fig-expt} 
$n_c=6$ and each codon is UUU that codes for the amino acid 
Phenylalanine (which is denoted either by its abbreviation 
{\it Phe} or the symbol {\bf F}). The coding sequence is preceeded
and followed by a start codon AUG and a stop codon UAA, respectively. 
A 5'-UTR preceeds the start codon; this UTR is required for assembling 
the ribosome and for stabilizing the pre-initiation complex. 
At the 3'-end, after the stop codon there is a 3'-UTR consisting of 
a sequence of $n_{nc}$ non-coding codons UUU 
($n_{nc}=4$ in the example of fig.\ref{fig-expt}); this region merely
ensures that the translation does not suffer from any ``edge effect'' 
when the ribosome approaches the 3'-end of the codong sequence. 
For such a poly-U mRNA sequence, aa-tRNA$^{Phe}$ is the cognate 
aa-tRNA. Translational error can be studied using this protocol 
if, in addition to cognate aa-tRNA$^{Phe}$, near-cognate 
aa-tRNA$^{Leu}$ is also supplied because the latter is cognate for 
the codon CUU which codes for {\it Leucine} (abbreviated {\bf L}).
An optical method, based on the labelling of the cognate, near-cognate 
and non-cognate tRNA molecules with dyes of different colors, has 
been suggested \cite{sharma11a} to test the validity of the expression 
derived for $f_{dwell}(t)$.


\subsection{Polysome: traffic-like collective phenomena}

\begin{figure}[ht]
\includegraphics[angle=0,width=0.8\columnwidth]{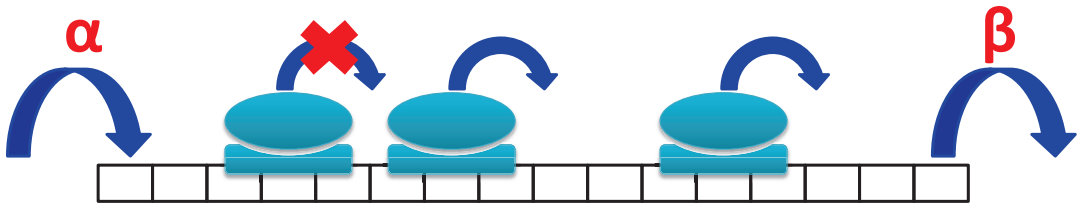}
\caption{ 
Ribosome traffic on mRNA track is shown schematically. 
The parameters $\alpha$ and $\beta$ denote the rates of translation 
initiation and termination, respectively. A ribosome can move forward 
by one codon if, and only if, the site immediately in front of it 
is not covered by another ribosome. The translation process is 
modelled more realistically by open boundary conditions, as shown 
here, than by periodic boundary conditions.
}
\label{rod_tasep2}
\end{figure}

The collective movement of many ribosomes on a single mRNA strand is 
shown schematically in fig.\ref{rod_tasep2}. It has superficial 
similarity with {\it single-lane uni-directional} vehicular traffic 
\cite{chowdhury00,schadschneider10} and is, therefore, sometimes 
referred to as ribosome traffic \cite{chowdhury05a}. 
The ribosomes bound simultaneously to a single mRNA transcript are 
the members of a polyribosome (or, simply, {\it polysome})
\cite{warner62,warner63,rich04,noll08}. 
Computer simulations of ribosome traffic have been carried out on a 
mRNA with a specially selected codon sequence near the start codon 
and allowing mRNA to decay at an optimum rate \cite{mitarai08}. 
In this case, the metabolic cost 
of mRNA breakdown is more than compensated by the simultaneous 
increase in translation efficiency because of reduced queing of 
the ribosomes. 

The polysome profiling technique \cite{arava03,mikamo05} provides the 
number of ribosomes bound to a mRNA, but not their individual positions 
where they remained ``frozen'' when translation was stopped by the 
experimental protocol. More detailed information on the numbers of
ribosomes associated with specified {\it segments} of a particular mRNA
can be obtained by using {\it ribosome density mapping} technique
\cite{arava05} which is based on site-specific cleavage of the mRNA 
transcript. However, the ribosomes are not expected to be uniformly 
distributed on the mRNA template. The detailed spatial distribution 
of the ribosomes on the mRNA template can be obtained by the most 
recent technique, called {\it ribosome profiling} 
\cite{ingolia10,guo10,ingolia11}.
This technique effectively provides a ``snapshot'' of the ongoing 
translation by the actively engaged ribosomes on the mRNA template. 
There are three major steps in this method:
(i) The ribosomes are first ``frozen'' at their instantaneous positions; 
(ii) the exposed parts of the mRNA transcripts (i.e., those segments 
not covered by any ribosome) are digested by RNase enzymes and, 
thereafter, the small ribosome ``footprints'' (segments protected 
by the ribosomes against the RNases) are collected separately; 
(iii) Finally, the ribosome-protected mRNA fragments thus collected 
are converted into DNA which are then sequenced. ``Aligning'' these 
footprints to the genome reveals the positions of the ribosomes at 
the instant when they were suddenly frozen.

Almost all the theoretical models of ribosome traffic represent the mRNA 
as a one-dimensional lattice where each of the $L$ sites corresponds to 
a single codon. Since an individual ribosome is much larger than a single 
codon, each ribosome is represented by a hard rod that can cover ${\ell}$ 
successive codons (${\ell} > 1$) simultaneously. Therefore, for the 
convenience of modeling, the mRNA template of $L$ codons can be represented 
by a one-dimensional lattice of $L+{\ell}-1$ sites where the first $L$ 
sites from the left represent the $L$ codons; the first and the $L$-th 
sites correspond to the start and stop codons, respectively. The position 
of a ribosome is denotes by the leftmost site of the lattice covered by it. 
Thus, a ribosome locatedat the $i$-th site covers all the ${\ell}$ sites 
from $i$ to $i+{\ell}-1$.

In this approach, ribosome traffic is treated as 
a problem of non-equilibrium statistical mechanics of a system of 
interacting ``self-driven'' hard rods on a one-dimensional lattice. 
Moreover, in these models the inter-ribosome interactions are 
captured through hard-core mutual exclusion principle: none of 
codons can be covered simultaneously by more than one ribosome. Thus, 
these models of ribosome traffic are essentially TASEP for hard rods: 
a ribosome hops forward, by one codon, with probability $q$ per unit 
time, if an only if the hop does not lead to any violation of the 
mutual exclusion principle. Since no backtracking of ribosome has been 
observed, total asymmetry of hopping of the ribosomes in the TASEP-type 
models is justified.  In TASEP-type models of ribosome traffic, 
all the details of the mechano-chemical cycle of a ribosome during 
the elongation stage is captured by a single parameter $q$. 
These models have been reviewed very recently from the 
perspective of statistical physics \cite{chou11}. Another recent 
review article has summarized mathematical models of ribosome 
movement and translation \cite{haar12}. Therefore, we'll not 
discuss these in any further detail here. However, it is worth 
emphasizing that, strictly speaking, a ribosome is neither just a 
particle nor merely a hard rod; it can exist in more than one 
``internal'' states that correspond to its various ``chemical'' states.  
Only a few attempts have been made in recent years to capture the 
detailed mechano-chemistry of individual ribosomes in the 
quantitative models of interacting ribosomes in trafic-like situations.
Chowdhury and collaborators have developed kinetic models that 
incorporate both the single-ribosome mechano-chemical kinetics as well 
as their steric interactions \cite{basu07,garai09,sharma11b}. 
The spatio-temporal organization of the ribosomes is characterized by 
the average number density and average flux in the steady state. Based 
on the quantities, Chowdhury and coworkers have plotted the dynamic 
phase diagrams for ribosome traffic in spaces spanned by experimentally 
accessible parameters. It may become possible to test these phase 
diagrams in near future using the ribosome profiling technique 
\cite{ingolia10,guo10,ingolia11} (or, its newer versions).

\section{Comparison with some other machines} 

\subsection{Comparison with non-ribosomal peptide synthesizing machines}

\begin{figure}[ht]
\includegraphics[angle=0,width=0.8\columnwidth]{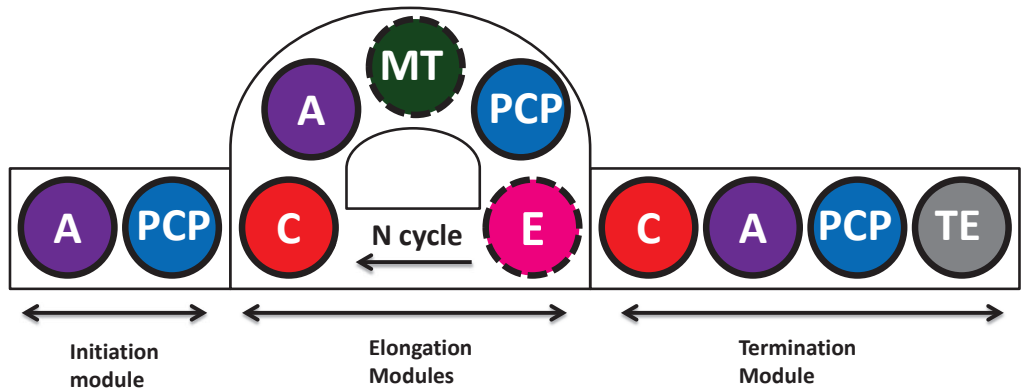}
\caption{Schematic representations of the chemical reaction involved in the synthesis of non ribosomal protein synthesis (adapted from \cite{challis04}). Condensation (C), adenylation (A) and peptidyl carrier (PCP) are the essential domains for the non ribosomal polypeptide elongation. The reaction starts when A domain activates the amino acid by consuming an ATP molecule and produces aminoacyl adenylate. This aminoacyl reacts with the thiol group, attached with PCP domain and the C domain allows the formation of peptide bond between the aminoacyl thioester group of the growing polypeptide chain of the last module with the aminoacyl thioester group of the current module.  Apart from these essential domains, some module also contain the Methyltransferase (MT) and epimerisation (E) domain which are responsible for extra enzymatic activities. MT is responsible for the methylation of the nitrogen of the amine whereas E domain cause the racemisation of the C$\alpha$ of the amino acid. The initiation module contain only A and PCP domain while the termination module includes one more extra TE domain, which cleaves the fully elongated polypeptide from PCP domain.}
\label{NRPS}
\end{figure}

\begin{figure}[ht]
\includegraphics[angle=0,width=0.8\columnwidth]{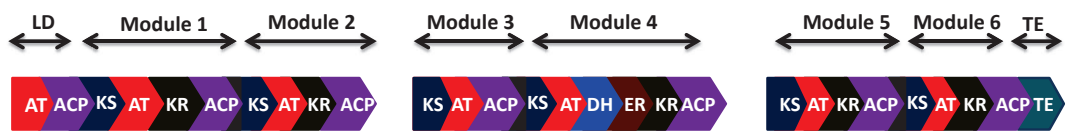}
\caption{Modules and domains involved in polyketide (6-deoxyerythronolide B synthase) biosynthesis (adapted from \cite{bedford96}).
 Ketosynthase (KS), acyl transferase (AT), and acyl carrier protein (ACP) are the minimal domains required in each module for the chain elongation whereas  ketoreductase (KR), dehydratase
(DH), and enoylreductase (ER) are the reductive domains and required for different type of condensations.}
\label{PKS}
\end{figure}

The monomeric subunits of non-ribosomal peptides and polyketides are 
amino acid and carboxylic acid, respectively. These two are the major 
families of natural products. Because of their importance in 
pharmaceutical and agrochemical industries, these natural products 
are the focus of attention of many leading labs \cite{wilkinson07}. 
A common feature of their synthesis 
\cite{crawford10,marahiel09,staunton01,fischbach06b,meier09} 
is that the non-ribosomal peptide synthetases (NRPS) (see fig.\ref{NRPS}) 
and the polyketide synthases (PKS) (see fig.\ref{PKS}) are both large 
proteins which contain repeated ``modules''. Each module, in turn, consists 
of more than one ``domain'' each of which has a specific activity. 
Each module performs one complete cycle of elongation of the polypeptide 
or polyketide chain. Thus, the length of the chain produced by such 
a machine depends on the number of modules on the machine.
The operations of the NRPS and PKS resemble that of an assembly-line 
\cite{chow04}. 

\begin{figure}[ht]
\includegraphics[angle=0,width=0.7\columnwidth]{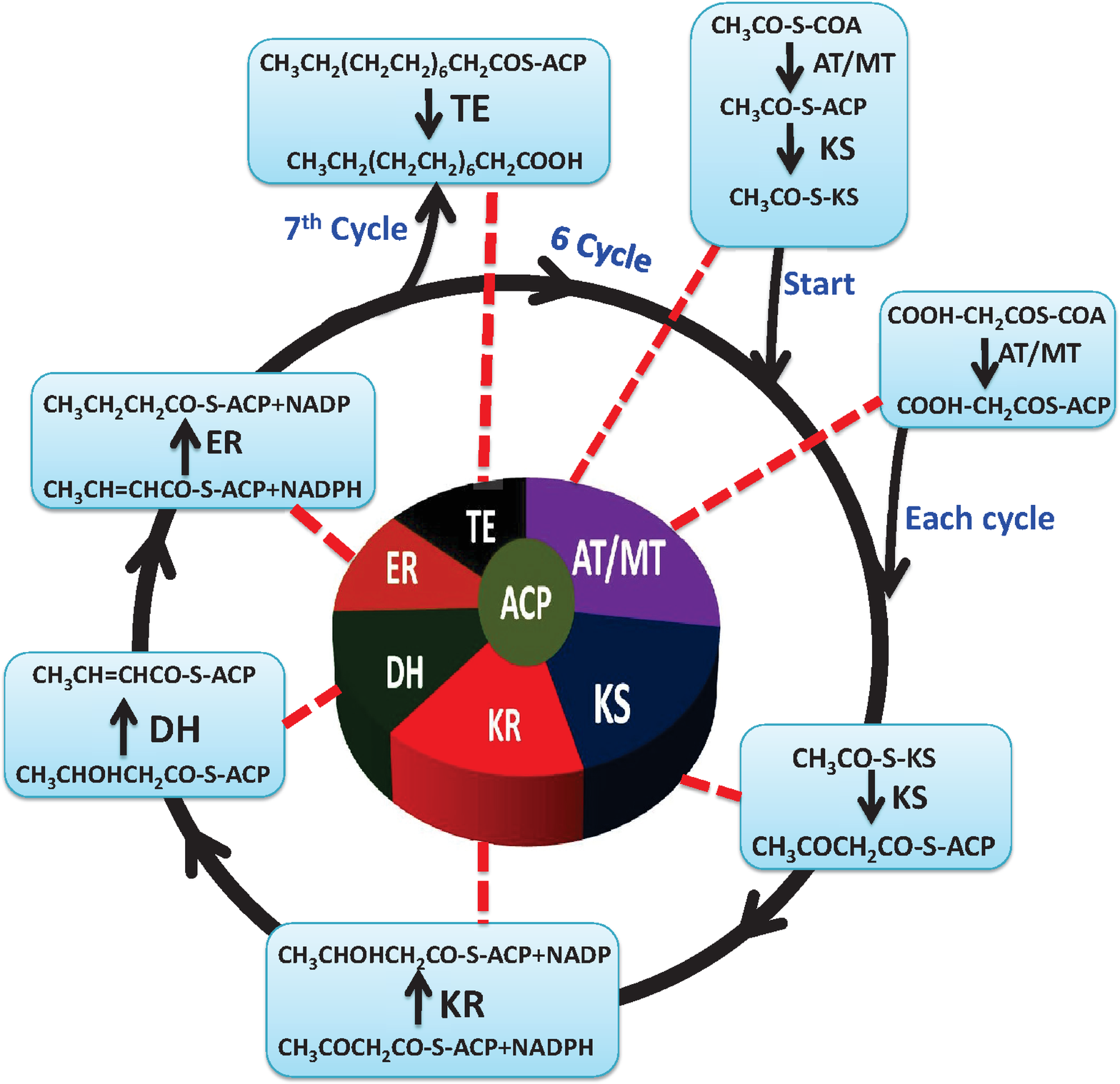}
\caption{Schematic representations of the chemical reactions involved in the synthesis of palmitate by FAS (adapted from \cite{chirala04}).  Reaction
starts when an acetyl group along with its coenzyme A is handed over to acyl carrier protein (ACP)
 by the action of acetyl/malonyl transacylase (MT/AT), from their it's transferred to  $\beta$-ketoacyl synthase (KS) and reacts with a malonyl group already attached with ACP. This reaction results in the condensation
of the acetyl group. Now this keto group is converted to the fully saturated carbon chain by the consecutive activities of $\beta$-ketoacyl reductase (KR),  $\beta$-hy­droxy-acyl dehydratase (DH) and enoyl reductase (ER).  After the 7th cycle of reaction ACP is released by thioesterase (TE).
}
\label{FAS}
\end{figure}

The modularity of the fatty acid synthase (FAS) \cite{khosla09} (see 
fig.\ref{FAS}) and its 
mechanism of chain elongation has similarities with those of PKS 
\cite{white05,tehlivets07,jakobsson06,leonard04,smith03}.
The structural investigations explore not only the structure at the 
level of single domains, but also on the intra-module connections of 
the domains as well as the inter-module connections \cite{koglin09}.

NRPS-mediated polypeptide synthesis does not use mRNA as a template. 
Instead, the identity and the order of the protein domains of the 
synthetase serve as the template \cite{fischbach06b,lautru04}. 
In each module of a NRPS one (or two) of the leading domains serve as 
``gate-keeper'' and specifies the identity of the monomer to be selected. 
Thus, in contrast to polynucleotide templates 
discussed in the preceeding sections, proteins serve as the templates 
for NRPS-mediated polymerization. 
The substrate specificity of NRPS and PKS raises questions which are 
similar to those addressed earlier in the context of ribosomal polypeptide 
synthesis, namely, the mechanisms of proofreading, error tolerance 
and the fidelity of the process \cite{khosla99,grunewald06}. 
Just like in ribosomal polypeptide 
synthesis, one can identify three stages, namely, initiation, chain 
elongation and termination also in the synthesis of non-ribosomal 
peptides. 
An approximate correspondence between a ribosome and  a NRPS may help 
in comparing the structure and function of these two types of machines 
\cite{dohren99,finking04}.
(i) In ribosomal synthesis, the aa-tRNA synthetase first selects the 
cognate amino acid and loads it onto the corresponding tRNA thereby 
charging it as an amino-acyl tRNA. 
Similarly, the A-domain of the NRPS selects the cognate amino acid 
and activates it as amino-acyl adenylate. 
(ii) The transfer of the amino-acid carrier, i.e., aatRNA, to the P 
site is assisted by the GTP hydrolysis catalyzed by the enzyme EF-Tu. 
The analogous process in the NRPS is the handing over of the activated 
amino acid to the peptidyl carrier protein (PCP) that transfers it 
to the C-domain. 
(iii) Just as the peptidyl transferase enzyme catalyzes the peptide 
bond formation on the ribosome, the C-domain of the NRPS catalyzes 
the peptide bond formation in the non-ribosomal pathway.  

\subsection{Comparison between polynucleotide polymerases and  
cytoskeletal motors} 
Let us compare these polymerase motors with the cytoskeletal motors. 
(i) Polymerase motors generate forces which are about 3 to 6 times 
stronger than that generated by cytoskeletal motors. 
(ii) The step size of a polymerase is about 0.34 nm whereas that 
of a kinesin is about 8 nm. 
(iii) The polymerase motors are slower than the cytoskeletal 
motors by two orders of magnitude. 
(iv) Natural nucleic acid tracks are intrinsically 
inhomogeneous because of the inhomogeneity of nucleotide sequences 
whereas the cytoskeletal tracks are 
homogeneous and exhibit perfect periodic order.

\subsection{Polymerases as ``tape-copying Turing machines'': dissipationless computation}

Template-directed polymerization has been analyzed in terms of the 
principles of information theory. The pioneering works were carried 
out by Wolkenshtein and Eliasevich \cite{wolkenshtein61} and by Davis 
\cite{davis65}. These authors calculated the lowering of entropy, 
i.e.,generation of information, in template-directed polymerization. 
These initial calculations were based on the simplifying assumption 
that each nucleotide addition is an event independent of that of 
the neighboring ones on the same template. In a recent work Arias-Gonzalez 
\cite{gonzalez12} has extended these theories by incorporating the 
effect of interactions between the neighboring nucleotides. 
However, Arias-Gonzalez \cite{gonzalez12} assumed an equilibrium pathway 
for the process which implies absence of dissipation. Obviously, this 
is not valid for any real template-directed polymerization process.  
Effects of the nonequilibrium conditions of template-directed 
polymerization processes on the resulting information transmission were 
investigated by Andrieux and Gaspard \cite{andrieux08,andrieux09}. 

A Turing machine carries out computation by repeating a cycle of 
logical operations. In each cycle it reads input information from a 
digital tape and, then, produces an output based on a set of rules.
The translocation of a RNA polymerase along its template resembles a 
Turing machine in the sense that it also moves along a digital tape 
(DNA), reads information from it, and produces an output as a result 
of its ``computation'' based on its ``rules''. However, the output, 
namely the RNA, is another digital tape. Therefore, a RNAP (and, 
similarly, a ribosome) can be regarded as a ``tape-copying Turing 
machine'' \cite{turing36} that polymerizes its output tape, instead 
of merely writing on a pre-synthesized tape \cite{bennett82}. 

Polymerization of RNA by RNAP has been analyzed also by Mooney et al. 
\cite{mooney98} from the perspective of information processing. 
Unlike the digital logic of a computer, decisions made by a polymerase 
are governed by competing rates and equilibria among alternative 
conformations and complexes. The changes in these conformations, 
i.e., the regulatory decisions made by a RNAP depend on two types of 
information input (to be distinguished from energy input): 
(i) intrinsic, and (ii) extrinsic. 
The intrinsic input include, for example, (a) the segment of the 
DNA template within the transcription bubble, (b) the nascent RNA, etc. 
whereas all the transcription factors are extrinsic inputs. Depending 
on the regulatory decisions made by the RNAP, it either elongates the 
RNA or pauses, or backtracks. 
Although futile cycles cause dissipation, these are essential for 
error correction. Dissipationless operation of these machines is 
possible only if every step is error-free which, in turn, is achievable 
in the vanishingly small speed, i.e., reversible limit \cite{bennett79}.
In other words, a polymerase is a Maxwell's demon that ``accumulates 
and stores'' information by creating an ordered sequence of nucleotides, 
as directed by the corresponding template \cite{klump09}.

\section{Summary and conclusion}

In this article we have reviewed some of the recent progress in understanding 
the common features of template-directed polymerization. We have discussed 
the structural features of the machines and the kinetic processes that they 
drive. Each faces conflicting requirements of speed and fidelity of 
polymerization. These tape-copying Turing machines hold promise for 
physical realization of dissipationless computation in future. We have also 
drawn attention to another class of modular machines which also carry out 
an altogether different type of template-directed polymerization. Such 
natural machines in living systems have already inspired designing of 
artificial ``molecular assembly line'' \cite{chow04}. We hope theoretical 
models would be developed for these modular machines in near future for 
making quantitative characterization of their operational mechanism.
Finally, fundamental questions on the constraints imposed by the ``local 
detailed-balance conditions'' on the rates of the various motor-driven 
processes \cite{seifert11} need to be addressed in the context of 
template-directed polymerization.


\noindent {\bf Acknowledgements}: One of the authors (DC) thanks Ashok 
Garai and Tripti Tripathi for many discussions on the topic of this 
review and for enjoyable collaborations. This work has been supported 
at IIT Kanpur by the Dr. Jag Mohan Garg Chair professorship (DC) and 
a CSIR fellowship (AKS). It has also been supported, in part, at the 
Ohio State University, Columbus, by the Mathematical Biosciences 
Institute and the National Science Foundation under grant DMS 0931642. 


\end{document}